\documentclass[cis]{ipart_v1}
\firstpage{1}
\Vol{}
\Issue{}
\Year{}
\usepackage{graphics}
\usepackage{amsmath} 
\usepackage{amssymb}   
\usepackage{amsthm}
\usepackage{paralist}
\allowdisplaybreaks
\newtheorem{theorem}{Theorem}[section]
\newtheorem{lemma}[theorem]{Lemma}
\newtheorem{definition}[theorem]{Definition}
\usepackage{enumitem}
\setlist{noitemsep}
\makeatletter
\newcommand{\inlineitem}[1][]{%
\ifnum\enit@type=\tw@
    {\descriptionlabel{#1}}
  \hspace{\labelsep}%
\else
  \ifnum\enit@type=\z@
       \refstepcounter{\@listctr}\fi
    \quad\@itemlabel\hspace{\labelsep}%
\fi}
\makeatother
\begin{document}
\setlength{\abovedisplayskip}{0pt}
\setlength{\belowdisplayskip}{0pt}
\setlength{\abovedisplayshortskip}{0pt}
\setlength{\belowdisplayshortskip}{0pt}

\title{Order Effects of Measurements in Multi-Agent Hypothesis Testing\textsuperscript{1}}

\author{Aneesh Raghavan$^\ast$, John S. Baras $^\ast$\blfootnote{(1) This paper is dedicated to Professor Tyrone Duncan on the occasion of his 80th birthday. John Baras would like to acknowledge the many discussions on these topics he had with Prof. Duncan since 1980. Aneesh Raghavan would like to thank Prof. Karl Henrik Johansson, KTH, Stockholm  for hosting him twice at KTH, Stockholm during which significant portion of this work was completed. \\Research supported by ARO grant W911NF-15-1-0646, by DARPA through ARO grant W911NF-14-1-0384 and by ONR grant N00014-17-1-2622.}}

\begin{abstract}
In multi-agent systems, agents observe data, and use them to make inferences and take actions. As a result sensing and control naturally interfere, more so from a real-time perspective. A natural consequence is that in multi-agent systems there are propositions based on the set of observed events that might not be simultaneously verifiable, which leads to the need for probability structures that allow such \textit{incompatible events}. We revisit the structure of events in a multi-agent system and we introduce the necessary new models that incorporate such incompatible events in the formalism. These models are essential for building non-commutative probability models, which are different than the classical models based on the Kolmogorov construction. From this perspective, we revisit the concepts of \textit{event-state-operation structure} and the needed \textit{relationship of incompatibility} from the literature and use them as a tool to study the needed new algebraic structure of the set of events.  We present an example from multi-agent hypothesis testing where the set of events does not form a Boolean algebra, but forms an ortholattice. A possible construction of a `noncommutative probability space', accounting for \textit{incompatible events} is discussed. We formulate and solve the binary hypothesis testing problem in the noncommutative probability space. We illustrate the occurrence of `order effects' in the multi-agent hypothesis testing problem by computing the minimum probability of error that can be achieved with different orders of measurements.
\end{abstract}

\maketitle

\section{Introduction}
In the study of stochastic multi-agent decision making problems it is often assumed that the joint distribution of the observations collected by the agents is known to all agents. In our previous work  \cite{raghavan2019binary}, we have discussed as to why such an assumption will not lead to \textit{truly} decentralized policies. Hence the joint distribution of measurements collected by agents in a multi-agent system might not be always available. When a probability space is to be constructed for an agent in the multi-agent system, the first step would be to enumerate the list of events / propositions that the agent can verify. We recall that in Kolomogorov's axioms for classical probability, it is assumed the set of events (associated with subsets of sets) form a Boolean algebra, a very specific algebraic structure. Hence, the existence of a classical probability space for formulating and solving decision-making problems imposes restrictions on the set of events, i.e., the set of verifiable propositions. By assuming that we can construct a classical probability space we assume that the set of events is a Boolean algebra. This assumption implies that all subsets of events are simultaneously verifiable. In multi-agent systems, agents collect observations and exchange information. In asynchronous multi-agent systems, the agents might not have a common notion of time. Propositions involving information from different agents might not be simultaneously verifiable as the information might not be simultaneously available, thus violating the structure of a Boolean algebra. Hence before we construct a classical probability space for an agent, we would first  have to verify that the set of events indeed form a Boolean algebra and cannot assume a priory that a classical probability space can be constructed for the agent. 

\noindent Our hypothesis is that the algebraic structure of the set of events need not be a Boolean algebra, it can be an orthomodular ortholattice. We present an example from multi-agent decision making supporting our hypothesis. This hypothesis is motivated from the observation that for an agent there could exist propositions which are not ``simultaneously verifiable" by the agent. Such events exist in quantum mechanical systems, which leads to the set of events forming an orthomodular ortholattice. The algebraic structure of the set of events in quantum mechanical systems have been well investigated in literature. One of the earliest papers in this direction, is \cite{birkhoff1936logic}. More recently, in \cite{hintikka2002quantum} the author argues that quantum logic is a fragment of independent friendly logic. Noncommuting observables are assumed to be mutually dependent variables. Independent friendly logic allows all possible patterns of dependence/ independence to be expressed among variables, which is not possible in first order logic \cite{baras2016multiagent}. Independent friendly logic violates the law of excluded middle ( every proposition, either in its positive or negative form is true). This violation stems from the fact that truth value for propositions is assigned by finding a winning strategy for a player in a suitable game. Since a winning strategy need not always exist, it is not always possible to assign truth values. In \cite{hintikka2002quantum}, the author argues that one can  find a suitable analogy between quantum logic and an extension of independent friendly logic. 

\noindent Our objective is to study multi-agent decision-making problems given ``data sets" or samples of (observation, decision) pairs generated from the multi-agent system. Our objective leads us to first study the algebraic structure of the set of events and then ``suitably" construct a probability space where the decision-making problems can be formulated and solved \cite{baras1987distributed, baras2003multiagent,baras2016multiagent}.  The problem that we consider is the binary hypothesis testing problem with three observers and a central coordinator. There are two possible states of nature, one of which is the true state of nature.  There are three observers collecting measurements (samples) that are statistically related to the true state of nature. The joint distribution of the measurements collected by the observers is unknown. Each observer knows the marginal distribution of the observations it alone collects. Each observer performs sequential hypothesis testing and arrives at a binary decision. The binary decision is then sent to a central coordinator. The objective of the central coordinator is to find its own belief about the true state of nature by treating the decision information that it receives as measurements. At the central coordinator a \textit{suitable} probability space is to be constructed for formulating and solving the hypothesis testing problem.

\noindent Our contributions are as follows. The set of events, i.e, the set of propositions that can be verified by the central coordinator is enumerated. We show that the set along with a suitable relation of implication and a unary operation of orthocomplmentation is not a Boolean algebra \cite{baras1987distributed, baras2003multiagent,baras2016multiagent}. To prove the same, we adopt the methodology developed in \cite{pool1968baer}. Hence the construction of a classical probability space is ruled out. We construct an \textit{event-state structure} (a generalization of measure spaces) for the central coordinator along the lines of the von Neumann Hilbert space model. We associate \textit{operations} (a generalization of conditional probability) \cite{baras1987distributed, baras2003multiagent,baras2016multiagent} with the event-state structure and construct a noncommutative probability space for the central coordinator. We consider the binary hypothesis testing problem in the non-commutative probability framework. We adopt the fomulation and solution methodology from our previous work \cite{baras1987distributed, baras2003multiagent,  baras2016multiagent, raghavan2019non, raghavan2019thesis}. We represent the different orders of measurements by different Positive Operator Valued Measures and solve the hypothesis testing problem \cite{baras1987distributed, baras2003multiagent,  baras2016multiagent, raghavan2019non, raghavan2019thesis}. For specific empirical local distributions at the three observers, we demonstrate that different orders of measurement can lead to different minimum probabilities of error at the central coordinator.

\noindent The paper is organized as follows. In the next section, section \ref{Algebraic structure of the set of events}, we present the methodology from \cite{pool1968baer} which we use to investigate the structure of the set of events. In section \ref{Example: multi-agent decision making}, we discuss a specific example from multi-agent decision making supporting our hypothesis. In section \ref{Binary hypothesis testing problem}, we discuss a hypothesis testing problem in a non commutative probability space, the probability space following the von Neumann Hilbert space model. 
\section{Algebraic structure of the set of events}\label{Algebraic structure of the set of events}
To keep this paper self contained, we introduce some definitions and identities from propositional calculus that have been mentioned in the literature \cite{birkhoff1936logic}.
\vspace{-\baselineskip}
\subsection{Introduction to propositional calculus} \label{Algebraic structure of the set of events Introduction to propositional calculus}
Let $\mathbb{E}$ be an experiment. Let $\mathbb{B}$ be the set of experimentally verifiable propositions, i.e.,  propositions to which we can assign truth value based on the outcome of the experiment $\mathbb{E}$. 

\noindent \textit{Example 1} \cite{specker1975logic}. Let the experiment $E$ be `observing the environment (surroundings)'. Suppose the set of propositions is $B$ =$\{$it is raining, it is snowing, it is warm, it is cold, the sun is  shining,  it is not raining, it is not snowing, it is not warm, it is not cold,  the sun is not shining$\}$. By performing the experiment(i.e., by observing the surroundings) one can assign truth value to each proposition, i.e., each proposition is either true or false. 

\noindent On the domain of propositions, we are given the relation of \textit{implication}($\leq$) which satisfies the following properties:
\begin{inparaitem}
\item[(i)] \textit{reflexive}: for any proposition $p_{1} \in \mathbb{B}$, $p_{1} \leq p_{1}$;
\item[(ii)]\textit{transitive}: for propositions $p_{1}, p_{2}$ and $p_{3}$ belonging to $\mathbb{B}$, if $p_{1} \leq p_{2}$ and $p_{2} \leq p_{3}$, then $p_{1} \leq p_{3}$. 
\end{inparaitem}
In example 1, `it is warm' $\leq$ `it is not snowing', `it is raining'  $\leq$ `it is not shining' and, `it is cold' $\leq$ `it is snowing' (this implication need not be true always). We can define the relation of \textit{co-testable} on the set of propositions as follows: two propositions are co-testable if and only if they can be assigned truth values simultaneously. This relation is reflexive, symmetric but is not transitive. When we verify the relation of implication between two propositions $p_{1}$ and $p_{2}$, we are simultaneously assigning truth value to both the propositions, i.e., we are assuming that the propositions are co-testable. If we impose the condition that the relation of implication between two propositions can be verified only when the propositions are co-testable, we lose the transitivity property of the relation of implication. The concept of simultaneous testability was introduced in \cite{specker1975logic}.   

\noindent The domain $\mathbb{B}$ and the relation implication , $\mathbb{L} = (\mathbb{B},\leq)$ form \textit{a partially ordered set}(POSET). The transitivity property of the relation of implication is essential for the construction of the partially ordered set. Hence, we assume every pair of propositions is co-testable.  We assume that the domain $\mathbb{B}$ includes the identically true proposition, denoted by $\mathbf{1}$, and the identically false proposition, denoted by $\mathbf{0}$. Both $L=(B, \leq)$ and $\hat{L}=(\hat{B} = B\cup \{\mathbf{0}, \mathbf{1}\}, \leq)$ are partially ordered sets. Using the relation of implication, we can define operations on the set $\mathbb{B}$. 
\vspace{-\baselineskip}
\begin{definition}
Let $\mathbb{L} =(\mathbb{B}, \leq)$  be a POSET. A proposition $p \in \mathbb{B}$ is said to be the conjunction (greatest lower bound or ``meet") of propositions $p_{1} \in\mathbb{B}$ and $p_{2} \in \mathbb{B}$ if $p \leq p_{1}$, $p \leq p_{2}$, and, for any other proposition $q \in \mathbb{B}$ such that $q \leq p_{1}$ and $q \leq p_{2}$, $q \leq p$. The conjunction of $p_{1}$ and $p_{2}$ is denoted by $p_{1} \wedge p_{2}$.
\end{definition} 
\vspace{-\baselineskip}
\begin{definition}
Let $\mathbb{L} =(\mathbb{B}, \leq)$  be a POSET. A proposition $p \in \mathbb{B} $ is said to be the disjunction (least upper bound or ``join") of propositions $p_{1} \in \mathbb{B}$ and $p_{2} \in \mathbb{B}$ if $p_{1} \leq p$, $p_{2} \leq p$, and, for any other proposition $q \in \mathbb{B}$ such that $p_{1} \leq q$ and $p_{2} \leq q$, $p \leq q$.  The disjunction of $p_{1}$ and $p_{2}$ is denoted by $p_{1} \vee p_{2}$.
\end{definition}
\vspace{-\baselineskip}
\begin{definition}
Let $\mathbb{L} =(\mathbb{B}, \leq)$  be a POSET. A proposition $p \in \mathbb{B}$ is said to be logically equivalent to proposition $q \in \mathbb{B}$ if $p \leq q$ and $q \leq p$. 
\end{definition}
\vspace{-\baselineskip}
\noindent In the example, the meet and join of the propositions are not included in $B$. We obtain the set $\bar{B}$, by taking the closure of the set $B$ with respect to the conjunction and disjunction operations. $\bar{L} = (\bar{B},\leq)$ is also a POSET.
\begin{definition}\label{Definition 1}
Let $\mathbb{L} =(\mathbb{B}, \leq)$  be a POSET with with $\mathbf{1}$and $\mathbf{0}$. A mapping $^{\prime}: \mathbb{B} \rightarrow \mathbb{B}$ is an orthocomplementation,(denoted by $^{\prime}$) provided it satisfies the following identities: for $p, p_{1},$ and $p_{2} \in \mathbb{B}$, 
\begin{enumerate}
\item[1.] $(p^{\prime})^\prime = p$,
\item[2.] $p \wedge p^{\prime} = \mathbf{0}$ and $p \vee p^{\prime} =\mathbf{1}$,
\item[3.] $p_{1} \leq p_{2}$ implies $p_{2}^{\prime} \leq p_{1}^{\prime}$.
\end{enumerate}
If $^{\prime}: \mathbb{B} \rightarrow \mathbb{B}$ is an orthocomplementation, the relation of orthogonality ($\perp$) is defined as $p_{1} \perp p_{2}$ if and only if $p_{1} \leq p^{\prime}_{2}$. 
\end{definition} 
\vspace{-\baselineskip}
\noindent The relation of orthogonality is not reflexive or transitive. From identity (3) of definition \ref{Definition 1}, it follows that the relation is indeed symmetric. From the definitions of the conjunction and disjunction operators and the identities, (1), (2), (3) of definition \ref{Definition 1}, the following result can be proven,
\begin{enumerate}
\item[4.] $(p_{1} \wedge p_{2})^{\prime} = p^{\prime}_{1} \vee p^{\prime}_{2}$ and $(p_{1} \vee p_{2})^{\prime} = p^{\prime}_{1} \wedge p^{\prime}_{2}$.
\end{enumerate} 
\vspace{-\baselineskip}
\begin{definition}\label{Definition 2}
A partially ordered set $\mathbb{L}=(\mathbb{B}, \leq)$ is said to be a lattice if: for every proposition $p_{1} \in \mathbb{B}$ and $p_{2} \in \mathbb{B}$, $p_{1}\wedge p_{2}$ and $p_{1} \vee p_{2} $ belong to $\mathbb{B}$.  
\end{definition}
\vspace{-\baselineskip}
\noindent From the above definition it follows that neither $L$ nor $\hat{L}$ are lattices but $\bar{L}$ is a lattice. The \textit{distributive} identity of propositional calculus can be stated as follows: for $p_{1}, p_{2}, p_{3} \in \mathbb{B}$, 
\begin{enumerate}
\item[5.]$p_{1} \vee (p_{2} \wedge p_{3}) = (p_{1} \vee p_{2}) \wedge (p_{1} \vee p_{3}) $ and \\ $p_{1} \wedge (p_{2} \vee p_{3}) = (p_{1} \wedge p_{2}) \vee (p_{1} \wedge p_{3}) $.
\end{enumerate}
A lattice which satisfies (2) of definition \ref{Definition 1} and (5) of definition \ref{Definition 2} is a Boolean algebra. In classical probability, the probability space consists of a sample space, a sigma algebra of subsets of the sample space and a probability measure on the sigma algebra. The sigma algebra along with set inclusion as the relation of implication, union of sets as the disjunction operation, and intersection of sets as conjunction operation is a Boolean algebra. Hence in classical probability we are defining measures over a Boolean algebra. The \textit{modular}     identity can be stated as follows:
\begin{enumerate}
\item[6.] If $p_{1} \leq p_{3}$, then $p_{1} \vee (p_{2} \wedge p_{3}) = (p_{1} \vee p_{2}) \wedge p_{3}$.
\end{enumerate}
The finite dimensional subspaces of a Hilbert space, along with subspace inclusion as the relation of implication, closed linear sum (instead of union of sets) as the disjunction operation, and set products (corresponding to intersection of sets) as conjunction operation satisfy the modular identity, but do not satisfy the distributive identity. Thus, if the propositions from the experiment along with the implication relation satisfy the modular identity, but not the distributive identity, they can be represented by the finite dimensional subspaces of a Hilbert space with the direct sum operation corresponding to the disjunction operation and the set product operation corresponding to the conjunction operation. In our study we consider the set of propositions as the propositions which describe the outcomes of experiments on multi-agent systems. They can be assigned truth values based on the outcome of the experiments. For propositions which arise from experiments on multi-agent systems, the relation of implication and unary operation of orthocomplementation are yet to be defined, but the properties and identities that they satisfy were discussed in this section. 
\subsection{Event state operation structure}\label{Algebraic structure of the set of eventsEvent state operation structure}
\subsubsection{Event-state structures}\label{Algebraic structure of the set of eventsEvent state operation structure Event-state structures}
We are interested in studying the structure of the set of experimentally verifiable propositions. We associate operations with the propositions (events as defined below) and measures on the set of propositions. From the properties of the operations and measures we infer the algebraic structure of the set of propositions. We follow the definitions mentioned in \cite{pool1968baer, baras1987distributed, baras2003multiagent,  baras2016multiagent}:
\vspace{-\baselineskip}
\begin{definition} 
An \textit{event state structure} is a triple $(\mathcal{E}, \mathbb{S}, \mathbb{P})$ where:
\begin{enumerate}
\item $\mathcal{E}$ is a set called the \textit{logic} of the event state structure and an element of $\mathcal{E}$ is called an event;
\item $\mathbb{S}$ is a set and an element of $\mathbb{S}$ is called a state;
\item $\mathbb{P}$ is a function $\mathbb{P}:\mathcal{E} \times \mathbb{S} \rightarrow [0,1] $ called the probability function and if $E \in \mathcal{E}$ and $\rho \in \mathbb{S}$ then $\mathbb{P}(E, \rho)$ is the probability of occurrence of event $E$ in state $\rho$;
\item If $E \in \mathcal{E}$, then the subsets $\mathbb{S}_{1}(E)$ and $\mathbb{S}_{0}(E)$ of  $\mathbb{S}$ are defined by $\mathbb{S}_{1}(E) = \{\rho \in \mathbb{S}: \mathbb{P}(E, \rho) =1\}, \; \mathbb{S}_{0}(E) = \{\rho \in \mathbb{S}: \mathbb{P}(E, \rho) =0\}$,
and if $\rho \in \mathbb{S}_{1}(E)$ ($\rho \in \mathbb{S}_{0}(E)$) then the event $E$ is said to occur (not occur) with certainty in the state $\rho$;
\item Axioms $I.1$ to $I.7$ below are satisfied. 
\end{enumerate}
\end{definition}
\noindent \textit{Axioms}:
\begin{enumerate}
\vspace{-0.2cm}
\item[I.1] If $E_{1},E_{2}$ belong to $\mathcal{E}$ and $\mathbb{S}_{1}(E_{1})= \mathbb{S}_{1}(E_{2})$ then $E_{1} = E_{2}$.
\item[I.2] There exists an event $\mathbf{1}$ such that $\mathbb{S}_{1}(\mathbf{1}) = \mathbb{S}$.
\item[I.3] If $E_{1}, E_{2}$ belong to $\mathcal{E}$ and $\mathbb{S}_{1}(E_{1}) \subset \mathbb{S}_{1}(E_{2}) $ then $\mathbb{S}_{0}(E_{2}) \subset \mathbb{S}_{0}(E_{1}) $. 
\item[I.4] If $E \in \mathcal{E}$ then there exists an event $E^\prime$ such that $\mathbb{S}_{1}(E)= \mathbb{S}_{0}(E^{\prime})$ and $\mathbb{S}_{0}(E) = \mathbb{S}_{1}(E^{\prime})$.
\item[I.5] If $E_{1}, E_{2}, ....$ are a sequence of events such that $\mathbb{S}_{1}(E_{i}) \subset \mathbb{S}_{0}(E_{j})$  for $ i \neq j$ then there exists a $E$ such that 
(a) $\mathbb{S}_{1}(E_{i}) \subset \mathbb{S}_{1}(E)$ for all i (b) if there exits $F$ such that $\mathbb{S}_{1}(E_{i}) \subset \mathbb{S}_{1}(F)$ for all i, then $\mathbb{S}_{1}(E) \subset \mathbb{S}_{1}(F)$, and (c) if $\rho \in \mathbb{S}$ then $\sum_{i}\mathbb{P}(E_{i}, \rho) =\mathbb{P}(E, \rho) $. 
\item[I.6] If $\rho_{1}, \rho_{2} \in \mathbb{S}$ such that $\mathbb{P}(E, \rho_{1}) = \mathbb{P}(E, \rho_{2})$ for every $E\in \mathcal{E}$ then $\rho_{1} = \rho_{2}$. 
\item [I.7] $\rho_{1}, \rho_{2},\ldots \in \mathbb{S}$, $t_{i} \in [0,1]$ and $ \sum_{i}t_{i} =1$ then exists an $\rho \in \mathbb{S}$ such that $\mathbb{P}(E, \rho)= \sum_{i}t_{i}P(E, \rho_{i})$ for all $E \in \mathcal{E}$. 
\vspace{-0.2cm}
\end{enumerate}
\noindent There are different interpretations that could be associated with the state, \cite{khrennikov2015quantum}. The state could refer to the physical state of the system. The state could be interpreted as a special (probabilistic) representation of information about the results of possible measurements on an ensemble of identically prepared systems. The second interpretation is appropriate given our context. An event may be identified with the occurrence or non-occurrence of a particular phenomenon pertaining to the multi-agent system. The event is associated with an observation procedure which interacts with the multi-agent system resulting in a yes or no corresponding to the occurrence or non-occurrence of the phenomenon. The interpretation of $\mathbb{P}(E, \rho)$ for $E \in \mathcal{E}$ and $\rho \in \mathbb{S}$ is as follows: we consider an ensemble of the systems such that the state is $\rho$. We determine the occurrence or non-occurrence of the event $E$ by executing the associated observation procedure associated with $E$ on each system in the ensemble \cite{baras2016multiagent}. If the ensemble is large enough then the frequency of occurrence of $E$ is close to $\mathbb{P}(E, \rho)$. Axiom I.1 states the condition for uniqueness of events. Axiom I.2 guarantees the existence of the certain event. Axiom I.4 guarantees the existence of the orthocomplement of any event. Axiom I.3 ensures that the third part of definition \ref{Definition 1} is satisfied. Axiom I.5 is equivalent to countable additivity of measures. Axiom I.6 states the condition for uniqueness of states. Axiom I.7 leads to $\sigma$ convexity of the probability function. 
\vspace{-\baselineskip}
\begin{definition}
If $(\mathcal{E}, \mathbb{S}, \mathbb{P})$ is an event state structure, then the relation of implication, $\leq$, is defined as follows: for $E_{1}, E_{2} \in \mathcal{E}$, $E_{1} \leq E_{2}$ if and only if $\mathbb{S}_{1}(E_{1}) \subseteq \mathbb{S}_{1}(E_{2})$. 
\end{definition}
\vspace{-\baselineskip}
\noindent The relation of implication is defined using the states and the probability function. Thus $E_{1}$ is said to imply $E_{2}$ if and only if the set of states for which $E_{1}$ occurs with certainty is a subset of the set of states for which $E_{2}$ occurs with certainty. Since the subset relation($\subseteq$) is reflexive and transitive, it follows that the implication relation is also reflexive and transitive. The antisymmetry property of the subset ($\subseteq$) relation and Axiom I.1 imply that the implication relation is also antisymmetric. Hence the relation of implication ($\leq$) is a partial ordering of $\mathcal{E}$. 
\vspace{-\baselineskip}
\begin{definition}
Let $(\mathcal{E}, \mathbb{S}, \mathbb{P})$ be an event state structure. Then the unique event $\mathbf{1} \in \mathcal{E}$ such that $\mathbb{S}_{1}(\mathbf{1}) = \mathbb{S}$ and $\mathbb{S}_{0}(\mathbf{1}) = \emptyset$ is the certain event. If $E$ belongs to $\mathcal{E}$, then the unique event $E^\prime \in \mathcal{E}$ such that $\mathbb{S}_{1}(E_{1})= \mathbb{S}_{0}(E^{\prime}_{1})$ and $\mathbb{S}_{0}(E_{1}) = \mathbb{S}_{1}(E^{\prime}_{1})$ is called the complement (negation) of $E$. The unique event $\mathbf{0} \in \mathcal{E}$ such that $\mathbb{S}_{1}(\mathbf{0}) = \emptyset$ and $\mathbb{S}_{0}(\mathbf{1}) = \mathbb{S} $ is the impossible event.
\end{definition}
\vspace{-\baselineskip}
\noindent Axiom I.2 implies the existence of the certain event and Axiom I.1 implies that the certain event is unique.  Further, the certain event is the greatest event corresponding to $\leq$, as $E \leq \mathbf{1}$, for all $E \in \mathcal{E}$. Axiom I.4 applied to the certain event yields the unique event $\mathbf{0}$ such that $\mathbb{S}_{1}(\mathbf{0})= \emptyset$, $\mathbb{S}_{0}(\mathbf{0}) = \mathbb{S}$ and $\mathbf{0} \leq E$ for all $E \in \mathcal{E}$.
\vspace{-\baselineskip}
\begin{theorem}
If $(\mathcal{E}, \mathbb{S}, \mathbb{P})$ is an event state structure, then \normalfont{\cite{baras1987distributed,baras2016multiagent,pool1968baer}}:
\begin{itemize}
\item $(\mathcal{E},\leq) $ is a POSET;
\item $\mathbf{1}$ and $\mathbf{0}$ are the greatest and least elements of the POSET,  $(\mathcal{E},\leq) $;
\item $E \rightarrow E^\prime$ is an orthocomplementation on $(\mathcal{E}, \leq)$;
\item If $E_{1},E_{2} \in \mathcal{E}$, the following are equivalent: (a) $E_{1} \leq E_{2}$, \\ (b) $\mathbb{S}_{1}(E_{1}) \subseteq \mathbb{S}_{1}(E_{2})$, (c) $\mathbb{S}_{0}(E_{2}) \subseteq \mathbb{S}_{0}(E_{1})$;
\item If $E_{1},E_{2} \in \mathcal{E}$, the following are equivalent: (a) $E_{1} \perp E_{2}$, \\(b) $\mathbb{S}_{1}(E_{1}) \subseteq \mathbb{S}_{0}(E_{2})$, (c) $E_{1} \leq E^{\prime}_{2}$;
\item If $E_{1}\in \mathcal{E}$, the following are equivalent: (a) $E_{1} = \mathbf{0}$, \\ (b) $\mathbb{S}_{1}(E_{1}) =\emptyset$, (c) $\mathbb{S}_{0}(E_{1}) =\mathbb{S}$.
\end{itemize}
\end{theorem}
\vspace{-\baselineskip}
\noindent For the proof of the above theorem we refer to \cite{pool1968baer}. 

\noindent \textit{Example 2} \cite{pool1968baer} We consider the classical probability model, the probability space constructed based on Kolomogorov's axioms. Let $\Omega$ be the sample space and $\mathcal{F}$ be a sigma algebra of subsets of $\Omega$. The relation of implication is defined as follows: $E_{1} \leq E_{2}$ if and only if $E_{1} \subseteq E_{2}$, where the relation $\subseteq$ is the set theoretic inclusion. $\mu: \mathcal{F} \rightarrow [0,1]$ is a probability measure if (a)$\mu(\emptyset) = 0$ and $\mu(\Omega) =1$ (b)if $\{E_{i}\}_{i \geq 1}$ is a sequence of pairwise orthogonal events, then $\mu(\cup_{i}E_{i}) = \sum_{i}\mu(E_{i})$. Let $\mathbb{S}$ be a collection of $\sigma$ convex, strongly order determining set of probability measures on $\mathcal{F}$. Let $\mathbb{P}(E, \rho) = \rho(E)$. Then $(\mathcal{F}, \mathbb{S}, \mathbb{P})$ is an event-state structure. The sample space $\Omega$ corresponds to the certain event (thus verifying Axiom I.2) and the $\emptyset$ corresponds to the impossible event. Axiom I.1 follows from the strong order determining property of the set $\mathbb{S}$. The orthocomplementation is given by $E^{\prime} = E^{c}$, where $^{c}$ demotes the set theoretic complement, satisfies Axiom I.4. Since $\mathcal{F}$ is a $\sigma$ algebra, the countable union of events in $\mathcal{F}$ also belongs to $\mathcal{F}$. This property of the $\sigma$ algebra along with countable additivity of the measures imply that Axiom I.5 is also satisfied. Axiom I.7 follows from the $\sigma$ convex property of the set $\mathbb{S}$.

\noindent \textit{Example 3} \cite{pool1968baer} Let $\mathcal{H}$ be a separable complex Hilbert space. Let $\mathbb{B}(\mathcal{H})$ denote the set of bounded linear operators which map from $\mathcal{H}$ to $\mathcal{H}$. Let $T^{\ast}$ denote the adjoint of $T \in \mathbb{B}(\mathcal{H})$. For $T \in \mathbb{B}(\mathcal{H})$, let $\mathcal{R}(T)= \{ u \in \mathcal{H}: u = T(v)\; \text{for some} \;v \in \mathcal{H}\}$ and $\mathcal{N}(T) = \{ v \in \mathcal{H}: T(v) = \theta \}$, where $\theta$ is the null vector of the Hilbert space. Let $\mathbb{B}^{+}_{s}(\mathcal{H})$ denote the set of Hermitian, positive semi-definite bounded linear operators. For the following definitions and results we refer to \cite{parthasarathy2012introduction}. Let $\mathbb{B}_{00}(\mathcal{H})$ denote the set of operators in $\mathbb{B}(\mathcal{H})$ which have finite rank. The set of compact operators $\mathbb{B}_{0}(\mathcal{H})$ is a closed subspace of $\mathbb{B}(\mathcal{H})$. The set $\mathbb{B}_{00}(\mathcal{H})$ is dense in $\mathbb{B}_{0}(\mathcal{H})$ with the operator norm.  Let $\{e_{i}\}_{i \geq 1}$ denote an orthonormal basis for $\mathcal{H}$ (since $\mathcal{H}$ is separable the orthonormal basis exists). For $T \in \mathbb{B}(\mathcal{H})$, the trace norm is defined as $||T||_{1} = \sum_{i}\langle |T|(e_{i}),e_{i}\rangle$, where $|T| =(T^{\ast}T)^{\frac{1}{2}}$ and $\langle\cdot,\cdot \rangle$ corresponds to inner product on the Hilbert space $\mathcal{H}$. The trace norm is independent of the choice of orthonormal basis. The set of trace class operators is the set of operators in $\mathbb{B}(\mathcal{H})$ which have finite trace norm, $\mathbb{B}_{1}(\mathcal{H}) =\{ T  \in \mathbb{B}(\mathcal{H}) : ||T||_{1} < \infty\}$. The set of trace class operators is a subspace of $\mathbb{B}(\mathcal{H})$. The vector space $\mathbb{B}_{1}(\mathcal{H})$ along with the trace norm $(||\cdot||_{1})$ is a  nonreflexive Banach Space. It can be shown that $||T|| \leq ||T||_{1}, T \in \mathbb{B}_{1}(\mathcal{H})$. $\mathbb{B}_{00}(\mathcal{H})$ is a dense subset of the Banach space $\mathbb{B}_{1}(\mathcal{H})$ with the trace norm. For $T \in \mathbb{B}_{1}(\mathcal{H})$, there exists $\{T_{n}\}_{n \geq 1} \subset \mathbb{B}_{00}(\mathcal{H}) $ such that $\{||T_{n} - T||_{1} \} \rightarrow 0$.  Since $||T|| \leq ||T||_{1}, \; T \in \mathbb{B}_{1}(\mathcal{H})$, $\{||T_{n} - T||\} \rightarrow 0$. When a sequence of compact operators converge to a bounded operator, that operator is also compact. Thus $T$ is compact. Hence $\mathbb{B}_{1}(\mathcal{H}) \subseteq \mathbb{B}_{0}(\mathcal{H})$, i.e., every trace class operator is compact. Let the closed (in norm topology) convex cone of hermitian, positive semidefinite trace class operators be denoted by $\mathcal{T}^{+}_{s}(\mathcal{H})$. Let $\mathbb{S} = \{T \in \mathcal{T}^{+}_{s}(\mathcal{H}):||T||_{1} =1 \}$. Let $\mathcal{P}(\mathcal{H})$ denote the set of all orthogonal projections onto $\mathcal{H}$, $\mathcal{P}(\mathcal{H}) = \{T \in \mathbb{B}(\mathcal{H}) : T \circ T = T , T^{\ast} = T \}$. Let $\mathbb{P}(E, \rho)$ for $E \in \mathcal{P}(\mathcal{H})$ and $\rho \in \mathbb{S}$ be defined as $\mathbb{P}(E, \rho) = Tr[\rho E] = \sum_{i}\langle \rho(E(e_{i})),e_{i}\rangle$. Then $(\mathcal{P}(\mathcal{H}),\mathbb{S},\mathbb{P})$ is an event state structure. The identity operator ($I_{\mathcal{H}}$) corresponds to the certain event and the null operator ($\Theta_{\mathcal{H}}$) corresponds to the impossible event. $I_{\mathcal{H}} \in \mathbb{B}(\mathcal{H})$ but does not belong to $\mathbb{B}_{1}(\mathcal{H})$. Axioms I.1, I.2 and I.3 can be verified. The orthocomplementation is given by $E^{\prime} = \mathbb{I}_{\mathcal{H}} - E$ which satisfies Axiom I.4. Axioms I.5 and I.6 can be verified. Since $\mathbb{S}$ is convex and the trace operator is linear, Axiom I.7 is also satisfied. $E_{1} \leq E_{2}$ if and only if $\{ \rho \in \mathbb{S} : Tr[\rho E_{1}] =1 \} \subseteq \{ \rho \in \mathbb{S} : Tr[\rho E_{2}] =1\}$ which is equivalent to stating that $E_{1}E_{2} = E_{1}$.  With this definition for the relation of implication, it can be shown that for $E_{1}, E_{2} \in \mathcal{P}(\mathcal{H})$, $E_{1} \wedge E_{2}$ is the projection onto the subspace $\mathcal{R}(E_{1}) \cap \mathcal{R}(E_{2})$ and $E_{1} \vee E_{2}$ is the projection onto the subspace $\mathcal{R}(E_{1}) \oplus \mathcal{R}(E_{2})$.
\subsubsection{Relation of compatibility}\label{Algebraic structure of the set of eventsEvent state operation structure Relation of compatability}
\begin{definition}
The relation of compatibility $\mathcal{C}$\normalfont{\cite{baras2016multiagent, pool1968baer}} is defined on the set of events, $\mathcal{E}$, as follows: for $E_{1}, E_{2} \in \mathcal{E}$, $E_{1} \mathcal{C} E_{2}$ if and only if there exists $F_{1}, F_{2}, F_{3} \in \mathcal{E}$ such that (a)$F_{1} \perp F_{2}$, (b)$F_{1} \perp F_{3}$, and $E_{1} = F_{1} \vee F_{3}$ and (c) $F_{2} \perp F_{3}$ and $E_{2} = F_{2} \vee F_{3}$.
\end{definition}
\vspace{-\baselineskip}
\noindent The relation $\mathcal{C}$ on $\mathcal{E}$ satisfies the following properties,\cite{pool1968baer}:
\begin{enumerate}
\item If $E_{1}, E_{2} \in \mathcal{E}$ and $E_{1} \leq E_{2}$ then $E_{1} \mathcal{C} E_{2}$;
\item If $E_{1}, E_{2} \in \mathcal{E}$ and $E_{1} \mathcal{C} E_{2}$ then (a) $E_{1} \mathcal{C} E^{\prime}_{2}$, (b) $E_{2} \mathcal{C} E_{1}$ (c) $E_{1} \wedge E_{2}$, and $E_{1} \vee E_{2}$ exist in $\mathcal{E}$;
\item If $E_{1}, E_{2}, E_{3} \in \mathcal{E}$, $E_{1} \mathcal{C} E_{2}$, $E_{2} \mathcal{C} E_{3}$, $E_{1} \mathcal{C} E_{3}$, and $(E_{1} \vee E_{3}) \wedge (E_{2} \vee E_{3})$ exists then $(E_{1} \wedge E_{2} ) \mathcal{C} E_{3}$ and $(E_{1} \wedge E_{2}) \vee E_{3} = \hspace{-0.1cm}(E_{1} \vee E_{3}) \wedge (E_{2} \vee E_{3})$. 
\end{enumerate}
The relation $\mathcal{C}$ is determined by the following property, \cite{pool1968baer}: for $E_{1} , E_{2} \in \mathcal{E}$, $E_{1} \mathcal{C} E_{2}$ if and only if there is Boolean sublogic $\mathcal{B} \subset \mathcal{E}$ such that $E_{1} , E_{2} \in \mathcal{B}$.
\vspace{-\baselineskip}
\begin{theorem} 
Let $(\mathcal{E}, \mathbb{S}, \mathbb{P})$ be an event state structure. If $E_{1} , E_{2} \in \mathcal{E}$ and there exists an $E_{3} \in \mathcal{E}$ such that $\mathbb{S}_{1}(E_{3}) = \mathbb{S}_{1}(E_{1}) \cap \mathbb{S}_{1}(E_{2}) $ then the conjunction of $E_{1} , E_{2}$ with respect to $\leq$ exists and is equal to $E_{3}$. 
\end{theorem}
\vspace{-\baselineskip}
\noindent For the proof of the above theorem we refer to \cite{pool1968baer}.
\subsubsection{Operations} \label{Algebraic structure of the set of events Event state operation structure Operations}
The concepts of conditional probability and conditional expectation are very important in classical probability theory. They enhance the utility of the theory and deepen the mathematical structure of the theory. They are extensively used in estimation, detection, filtering and control. Conditional probability is defined as a measure on a restricted sample space, with the `observed event' leading to the restriction. Conditional expectation of a random variable given a $\sigma$ algebra is a random variable which is measurable with respect to the $\sigma$ algebra and its expectation is equal to the expectation of the original random variable over the sets of the $\sigma$ algebra. Our goal is to obtain concepts analogous to conditional probability and conditional expectation for general event-state structures. Conditional probability can be viewed as a map from a probability measure to a probability measure restricted to the observed event. Since states in the event-state structure are ``analogous" to probability measures in classical probability, we first define maps from the set of states to the set of states and its associated properties. 
\vspace{-\baselineskip}
\begin{definition}
Let $(\mathcal{E}, \mathbb{S}, \mathbb{P})$ be an event state structure. 
\begin{enumerate}
\item Let $\mathbb{O}$ denote the set of all maps $T: \mathbb{D}_{T} \rightarrow \mathbb{R}_{T}$ with domain $\mathbb{D}_{T} \subset \mathbb{S} $ and range $\mathbb{R}_{T} \subset \mathbb{S} $. If $T \in \mathbb{O}$ and $\rho \in \mathbb{S}$ then $T(\rho)$ denotes the image of $\rho$ under $T$. 
\item For $T_{1}, T_{2} \in \mathbb{O}$, $T_{1} = T_{2}$ if and only if $\mathbb{D}_{T_{1}} = \mathbb{D}_{T_{2}}$ and \\ $T_{1}(\rho) = T_{2}(\rho),\; \forall \rho \in \mathbb{D}_{T_{1}}$.
\item \textit{0}$: \mathbb{D}_{\text{\textit{0}}} \rightarrow \mathbb{R}_{\text{\textit{0}}}$ is defined by $\mathbb{D}_{\text{\textit{0}}} = \emptyset$.
\item $\text{\textit{1}}: \mathbb{D}_{\text{\textit{1}}} \rightarrow \mathbb{R}_{\text{\textit{1}}}$ is defined by $\mathbb{D}_{\text{\textit{1}}} = \mathbb{S}$ and $\text{\textit{1}}(\rho) = \rho , \;\forall \rho \in \mathbb{S}$.
\item If  $T_{1}, T_{2} \in \mathbb{O}$, then $T_{1} \circ T_{2}: \mathbb{D}_{T_{1} \circ T_{2}} \rightarrow \mathbb{R}_{T_{1} \circ T_{2}}$ is defined by \\ $\mathbb{D}_{T_{1} \circ T_{2}} = \{\rho \in \mathbb{D}_{T_{2}}: T_{2}(\rho) \in \mathbb{D}_{T_{1}}\}$, $T_{1} \circ T_{2}(\rho) = T_{1}(T_{2}(\rho)) \; \hspace{-2pt} \forall \rho \in \mathbb{D}_{T_{1} \circ T_{2}}.$
\end{enumerate}
\end{definition}
\vspace{-\baselineskip}
\noindent In order to predict the result when consecutive experiments are performed on a system, it is essential to define the composition of maps. The state obtained by applying the composition of maps $T_{1}$ and $T_{2}$ to a state $\rho$, denoted by ($T_{1}\circ T_{2}(\rho)$), is the state obtained by applying the map $T_{2}$ first to $\rho$ and then applying $T_{1}$ to $T_{2}(\rho)$. We impose an axiomatic framework on the set of maps $(\mathbb{O})$ resulting in ``operations" which can be associated with events from the  experiment.  
\vspace{-\baselineskip}
\begin{definition}
An event-state-operation structure \cite{baras1987distributed, baras2003multiagent,  baras2016multiagent, pool1968baer}is a 4-tuple \\ $(\mathcal{E}, \mathbb{S}, \mathbb{P},\mathbb{T})$ where $(\mathcal{E}, \mathbb{S}, \mathbb{P})$ is an event-state structure and $\mathbb{T}$ is mapping \\ $\mathbb{T}: \mathcal{E} \rightarrow \mathbb{O}(\mathbb{T}: E \rightarrow T_{E})$ which satisfies Axioms II.1 to II.7 below.
\end{definition}
\vspace{-\baselineskip}
\noindent If $E \in \mathcal{E}$, then $T_{E}$ is called the operation \cite{baras1987distributed, baras2003multiagent,  baras2016multiagent, pool1968baer}  corresponding to the event $E$. If $E \in \mathcal{E}$ and $\rho \in \mathbb{D}_{T_{E}}$, then $T_{E}(\rho)$ is called the state conditioned on the event $E$ and state $\rho$. If $E_{1} \in \mathcal{E}$, then $\mathbb{P}(E_{1},T_{E}(\rho) )$ is the probability of $E_{1}$ conditioned on the event $E$ and state $\rho$. Let $\mathbb{O}_{T}$ denote the subset of $\mathbb{O}$ defined by $\mathbb{O}_{T} = \{T \in \mathbb{O}: T = T_{E_{1}} \circ T_{E_{2}} \ldots  \circ T_{E_{n}}; E_{1}, E_{2}, \ldots, E_{n} \in \mathcal{E}\}$. An element of $\mathbb{O}_{\mathbb{T}}$ is called an operation. 

\noindent \textit{Axioms:}
\begin{enumerate}
\item[II.1]If $E\in\mathcal{E}$, then the domain $\mathbb{D}_{T_{E}}$ of $T_{E}$ coincides with the set \\ $\mathbb{D}_{E} = \{\rho \in \mathbb{S}: \mathbb{P}(E,\rho) \neq 0 \}$.
\item[II.2]If $E\in\mathcal{E}$, $\rho \in \mathbb{D}_{E}$ and $\mathbb{P}(E, \rho)=1$ then $T_{E}(\rho)=\rho$.
\item[II.3]If $E \in \mathcal{E}$ and $\rho \in \mathbb{D}_{E}$, then $\mathbb{P}(E,T_{E}(\rho)) =1$.
\item[II.4]If $E_{1},E_{2}, \ldots, E_{n}$, $F_{1},F_{2}, \ldots, F_{n}$ are subsets of $\mathcal{E}$, and \\$T_{E_{1}} \circ T_{E_{2}} \circ \ldots \circ T_{E_{n}} = T_{F_{1}} \circ T_{F_{2}} \circ \ldots \circ T_{F_{n}}$ then \\ $T_{E_{n}} \circ T_{E_{n-1}} \circ \ldots \circ T_{E_{1}} = T_{F_{n}} \circ T_{F_{n-1}}\circ \ldots \circ T_{F_{1}}$.
\item[II.5]If $T \in \mathbb{O}_{T}$, then $\exists$ a $E_{T}$ such that  $\mathbb{S}_{1}(E_{T}) = \{\rho \in \mathbb{S}: \rho \notin \mathbb{D}_{T}\}$.
\item[II.6]If $E_{1}, E_{2} \in \mathcal{E}$, $E_{2} \leq E_{1}$ and $\rho \in \mathbb{D}_{E_{1}}$, then $\mathbb{P}(E_{2},T_{E_{1}}(\rho)) = \frac{\mathbb{P}(E_{2},\rho)}{\mathbb{P}(E_{1},\rho)} $.
\item[II.7]If $E_{1}, E_{2} \in \mathcal{E}$, $E_{1} \mathcal{C} E_{2}$ and $\rho \in \mathbb{D}_{E_{1}}$ then \\ $\mathbb{P}(E_{2},T_{E_{1}}(\rho)) = \mathbb{P}(E_{1} \wedge E_{2},T_{E_{1}}(\rho))$.
\end{enumerate}
\noindent \textit{Example 2} Operations for the classical probability space: The event state structure is $(\mathcal{F}, \mathbb{S}, \mathbb{P})$. For $E \in \mathcal{F}$, the operation is defined as  follows:
\begin{align*}
(T_{E}(\mu))(F) = \frac{\mu( E \cap F)}{\mu(E)}
\end{align*} 
The domain of $T_{E}$ is $\{\mu : \mu(E)\neq 0\}$, satisfying Axiom II.1. Axioms II.2, and II.3 can be verified. For Axiom II.4, it is given that $\frac{\mu(E_{1} \cap E_{2} \cap \ldots \cap E_{n} \cap G )}{\mu(E_{1} \cap E_{2} \cap \ldots \cap E_{n})} = \frac{\mu(F_{1} \cap F_{2} \cap \ldots \cap F_{n} \cap G )}{\mu(F_{1} \cap F_{2} \cap \ldots \cap F_{n})}$ for all $\mu$ in domain and $G \in \mathcal{F}$. Since the $\cap$ operation is commuting, it follows that $\frac{\mu(E_{n} \cap E_{n-1} \cap \ldots \cap E_{1} \cap G )}{\mu(E_{n} \cap E_{n-1} \cap \ldots \cap E_{1})} = \frac{\mu(F_{n} \cap F_{n-1} \cap \ldots \cap F_{1} \cap G )}{\mu(F_{n} \cap F_{n-1} \cap \ldots \cap F_{1})}$ for all $\mu$ in domain and $G \in \mathcal{F}$. For Axiom II.5, let $T_{E} = T_{E_{1}} \circ T_{E_{2}} \circ \ldots T_{E_{n}}$. The domain of $T_{E}$ is $\{\mu : \mu(E_{1} \cap E_{2} \cap \ldots \cap E_{n}) \neq 0\}$. The states which do not belong to the domain are: $\{\mu : \mu(E_{1} \cap E_{2} \cap \ldots \cap E_{n}) = 0\}$. 

\noindent Let $F=(E_{1} \cap E_{2} \cap \ldots \cap E_{n})^{c}$, that is the set theoretic complement of \\ $E_{1} \cap E_{2} \cap \ldots \cap E_{n}$. $F \in \mathcal{F}$ as $\mathcal{F}$ is a $\sigma$ algebra. \\ $\{\mu : \mu(E_{1} \cap E_{2} \cap \ldots \cap E_{n}) = 0\} = \{\mu: \mu(F) = 1\}$. Hence there exists a \\ unique event satisfying Axiom II.5. Axioms II.6 and II.7 can be verified. 

\noindent \textit{Example 3} Operations for the von Neumann Hilbert space model: given an event $E$, the operation corresponding to event $E$ is defined as:
\begin{align*}
T_{E}(\rho) = \frac{E \rho E}{Tr[\rho E]}
\end{align*}
The domain of $T_{E}$ is $\{\rho : Tr[\rho E] \neq 0\}$, satisfying Axiom II.1. Axioms II.2, and II.3 can be verified. For the verification of Axioms II.4 and II.5 we refer to sections \ref{Verification of axioms II.4 and II.5 Axiom II.4} and \ref{Verification of axioms II.4 and II.5 Axiom II.5}. Axioms II.6 and II.7 can be verified. We note that in this von Neumann Hilbert space model, the orthocomplementation corresponds to orthogonal complement of subspaces and not the set theoretic complement. This concept has been discussed in \cite{hintikka2002quantum}.  
\vspace{-\baselineskip}
\begin{definition}
Let $(\mathcal{E}, \mathbb{S}, \mathbb{P},\mathbb{T}))$ be an event-state-operation structure. The mapping $^{\ast}: \mathbb{O}_{\mathbb{T}} \rightarrow  \mathbb{O}_{\mathbb{T}}$ is defined as: if $T \in  \mathbb{O}_{\mathbb{T}}$, there exists $E_{1}, E_{2}, \ldots, \\ E_{n}$ such that $T= T_{E_{1}} \circ T_{E_{2}} \ldots  \circ T_{E_{n}}$, then $T^{\ast} = T_{E_{n}} \circ T_{E_{n-1}} \ldots  \circ T_{E_{1}}$.
\end{definition}
\vspace{-\baselineskip}
\noindent Axiom II.4 ensures that even if there are two sequences of operations which result in the same operation, i.e., for $T \in \mathbb{O}_{\mathbb{T}}$, $ \exists E_{1},E_{2}, \ldots, E_{n}$, $F_{1},F_{2}, \ldots, F_{n}$ subsets of $\mathcal{E}$ such that $ T= T_{E_{1}} \circ T_{E_{2}} \circ \ldots \circ T_{E_{n}} = T_{F_{1}} \circ T_{F_{2}} \circ \ldots \circ  T_{F_{n}}$, then the involution is unique as $T_{E_{n}} \circ T_{E_{n-1}} \circ \ldots \circ  T_{E_{1}} = T_{F_{n}} \circ T_{F_{n-1}}\circ \ldots \circ T_{F_{1}}$.
\vspace{-\baselineskip}
\begin{theorem}
If $(\mathcal{E}, \mathbb{S}, \mathbb{P},\mathbb{T}))$ is an event-state-operation structure, then $\mathbb{O}_{\mathbb{T}} $ is a subsemigroup of $\mathbb{O}$. Further, 
\begin{enumerate}
\item $T_{\mathbf{1}} = \text{\textit{1}}$ and $T_{\mathbf{0}} = \text{\textit{0}}$,
\item If $E \in\mathcal{E}$, then $T_{E} \circ T_{E} = T_{E}$, i.e., $T_{E}$ is a projection and the range of $T_{E} =  \mathbb{S}_{1}(E)$,
\item $^{\ast}: \mathbb{O}_{\mathbb{T}} \rightarrow \mathbb{O}_{\mathbb{T}}$ is the unique mapping such that 
\begin{enumerate}
\item $^{\ast}$ is an involution on the semigroup $(\mathbb{O}, \circ)$,
\item $(T_{E})^{\ast} = T_{E}$ for all $E \in \mathcal{E}$, and 
\item if $E_{1}, E_{2} \in \mathcal{E}$, then the following properties are equivalent: \\(i) $E_{1} \leq E_{2}$; \\(ii) $\mathbb{S}_{1}(E_{1})\subseteq \mathbb{S}_{1}(E_{2})$; \\ (iii) $\mathbb{S}_{0}(E_{2}) \subseteq \mathbb{S}_{0}(E_{1})$; \\ (iv) $T_{E_{1}} \circ T_{E_{2}} = T_{E_{1}}$; \\ (v) $T_{E_{2}} \circ T_{E_{1}} = T_{E_{1}}$.
\end{enumerate}
\end{enumerate}
\end{theorem}
\vspace{-\baselineskip}
\noindent For the proof we refer to \cite{pool1968baer}.
\noindent The theorem asserts that $(\mathbb{O}_{\mathbb{T}},\circ, ^{\ast})$ is an involution semigroup \cite{baras2003multiagent,  baras2016multiagent, pool1968baer} such that:
\begin{enumerate}
\item For each $E \in \mathcal{E}$, $T_{E}$ is a projection, that is $T_{E}$ belongs to the set $P(\mathbb{O}_{\mathbb{T}}) = \{T \in \mathbb{O}_{\mathbb{T}}	: T \circ T = T^{\ast} = T\}$.
\item $E\in \mathcal{E} \rightarrow T_{E} \in P(\mathbb{O}_{\mathbb{T}})$ is a order preserving map of $(\mathcal{E},\leq)$ into \\ $(P(\mathbb{O}_{\mathbb{T}}),\leq)$ where $T_{E} \leq T_{F}$ means $T_{E} \circ T_{F} = T_{E}$ for $T_{E},T_{F} \in P(\mathbb{O}_{\mathbb{T}})$.
\end{enumerate}	
\begin{definition}
If $(\mathcal{E}, \mathbb{S}, \mathbb{P},\mathbb{T}))$ is an event state operation structure then the mapping $^{\prime}: \mathbb{O}_{\mathbb{T}}\rightarrow P(\mathbb{O}_{\mathbb{T}})$ is defined as follows: for $T \in \mathbb{O}_{\mathbb{T}}$, $T^{\prime} = T_{E_{T}}$ where $E_{T} \in \mathcal{E}$ is the unique element of $\mathcal{E}$ such that \\ $\mathbb{S}_{1}(E_{T}) = \{\rho \in \mathbb{S}: \rho \notin \mathbb{D}_{T}\}$.
\end{definition}
\vspace{-\baselineskip}
\noindent Axiom II.5 ensures the existence of an event as required by the above definition. Uniqueness of the event follows from Axiom I.1. Axioms II.4 and II.5 were included to ensure that the involution and orthocomplementation operations can be defined on the set of operations. These operations are needed in order to construct a specific kind of semigroup, the Baer$ ^{\ast}$-semigroup, on the set of operations \cite{baras2003multiagent,  baras2016multiagent, pool1968baer}. This additional structure helps us find equivalence between compatibility of events and the commutativity of their corresponding operations. 
\vspace{-\baselineskip}
\begin{definition}
A Baer$ ^{\ast}$-semigroup $(S,\circ,^{\ast},^{\prime})$ is an involution semigroup $(S,\circ,^{\ast})$ with a zero $\text{\textit{0}}$ and a mapping $^{\prime}: S \rightarrow P(S)$ such that if $T \in S$ then $\{U \in S : T \circ U = \text{\textit{0}}\} = \{U \in S: U = T^{\prime} \circ V, \; \text{for some} \; V \in S\}$. If $(S,\circ,^{\ast},^{\prime})$ \\ is Baer$ ^{\ast}$-semigroup, then an element of $P^{\prime}(S) = \{T \in S: (T^{\prime})^{\prime} = T \}$ is called as closed projection. 
\end{definition}
\vspace{-\baselineskip}
\vspace{-\baselineskip}	
\begin{theorem}\label{Theorem-1}
Let $(S,\circ,^{\ast},^{\prime})$ be a Baer$ ^{\ast}$-semigroup. Then,
\begin{enumerate}
\item $P^{\prime}(S) = \{T \in S: (T^{\prime})^{\prime} = T \} = \{T^{\prime}; T \in S\}$.
\item If $T \in P^{\prime}(S)$, then $T^{\prime} \in P^{\prime}(S)$.
\item $(P^{\prime}(S),\leq, ^{\prime})$ is an orthomodular lattice where $\leq$ is the relation $\leq$ on $P(S)$ restricted to $P^{\prime}(S)$ and $^{\prime}$ is the restriction of $^{\prime}: S \rightarrow P(S)$ to $P^{\prime}(S)$. If $T_{1}, T_{2} \in P^{\prime}(S)$, then $T_{1} \wedge T_{2} = (T^{\prime}_{1} \circ T_{2})^{\prime}  \circ T_{2}$.
\item If $T,U \in P^{\prime}(S)$ then the following are equivalent:\\ (i) there exists $T_{0}, U_{0}, V_{0} \in P^{\prime}(S)$  such that $T_{0} \perp U_{0}$, $T_{0} \perp V_{0}$, $U_{0} \perp V_{0} $, $T = T_{0} \vee V_{0}$ and  $U = U_{0} \vee V_{0}$; \\ (ii) $T \circ U = U \circ T$. If $T \circ U = U \circ T$ then $T \wedge U = T \circ U$. 
\end{enumerate}
\end{theorem}
\vspace{-\baselineskip}
\noindent For the proof of the above theorem we refer to \cite{foulis1960baer}. From the axioms associated with operations, we conclude that $(\mathbb{O}_{\mathbb{T}}, \circ, ^{\ast}, ^{\prime})$ is a Baer$ ^{\ast}$-semigroup. Let $P^{\prime}(\mathbb{O}_{\mathbb{T}}) = \{T \in \mathbb{O}_{\mathbb{T}}: (T^{\prime})^{\prime} = T\}$. From the above theorem it follows that, $(P^{\prime}(\mathbb{O}_{\mathbb{T}}), \leq, ^{\prime})$ is an orthomodular ortholattice.

\noindent Commutative Baer$ ^{\ast}$-semigroup for \textit{Example 2}: Let $\mathbb{O}$ denote the set of \\ all maps from $\mathbb{S}$ to $\mathbb{S}$. Let $\mathbb{O}_{\mathbb{T}} =\{T \in \mathbb{O}: (T(\mu))(F) = \frac{\mu(E_{1} \cap E_{2} \cap \ldots \cap E_{n} \cap F )}{\mu(E_{1} \cap E_{2} \cap \ldots \cap E_{n})}, \\ E_{1}, E_{2}, \ldots E_{n} \in \mathcal{F}\}$. Since the axioms associated with involution and orthocomplmentation are satisfied, $(\mathbb{O}_{\mathbb{T}}, \circ, ^{\ast}, ^{\prime})$ forms a Baer$ ^{\ast}$-semigroup. Since the \\ set theoretic intersection operation ($\cap$) is commutative$(E_{1} \cap E_{2} = E_{2} \cap E_{1}, \\ E_{1} ,  E_{2}  \in \mathcal{F})$ the composition operation is commutative, i.e, $T_{1}\circ T_{2} = \\ T_{2} \circ T_{1},  T_{1},  T_{2}  \in \mathbb{O}_{\mathbb{T}}$. Thus   $(\mathbb{O}_{\mathbb{T}}, \circ, ^{\ast}, ^{\prime})$ is a commutative Baer$ ^{\ast}$-semigroup.

\noindent Noncommutative Baer$ ^{\ast}$-semigroup for \textit{Example 3}: First we note $(\mathbb{B}(\mathcal{H}), \circ)$ is a semigroup. The usual operator adjoint, $T \rightarrow T^{\ast}$ is an involution for  $(\mathbb{B}(\mathcal{H}), \circ)$. Let $\mathbb{B}_{O}(\mathcal{H}) = \{ T \in \mathbb{B}(\mathcal{H}): T = P_{1} \circ P_{2} \circ \ldots P_{n}, \{P_{i}\}^{n}_{i=1} \subset \mathcal{P}(\mathcal{H})\}$. It is clear that $(\mathbb{B}_{O}(\mathcal{H}), \circ, ^{\ast})$ is an involutive semigroup. For $T \in \mathbb{B}_{O}(\mathcal{H})$, the orthocomplementation of $T$ is the projection corresponding to the unique event satisfying Axiom II.5. Hence $(\mathbb{B}_{O}(\mathcal{H}), \circ, ^{\ast}, ^{\prime})$ is a Baer$ ^{\ast}$-semigroup. The semigroup is noncommutative as the composition of projections (multiplication of projections) is noncommutative. Let $\mathbb{O}$ denote the set of all maps from $\mathbb{S}$ to $\mathbb{S}$. Let $\mathbb{O}_{\mathbb{T}} =\{T \in \mathbb{O}: T(\rho) = \frac{ \prod^{n}_{i=1}E_{i}\rho \prod^{1}_{i=n}E_{i}}{Tr[\prod^{n}_{i=1}E_{i}\rho \prod^{1}_{i=n}E_{i}]}, E_{1}, E_{2}, \ldots E_{n} \in \mathcal{E}\}$. Every $ V \in \mathbb{B}_{O}(\mathcal{H})$, corresponds to a unique operation $T \in \mathbb{O}_{\mathbb{T}}$ and for every $T \in \mathbb{O}_{\mathbb{T}}$, there exists unique $V$ such that $T(\rho) = \frac{V^{\ast}\rho V}{Tr[V^{\ast}\rho V]} \forall\rho \in \mathbb{D}_{T}$. Thus $(\mathbb{O}_{\mathbb{T}}, \circ, ^{\ast}, ^{\prime})$ is also a noncommutative Baer$^{\ast}$ semigroup. \\$P^{\prime}(\mathbb{O}_{\mathbb{T}}) = \{T : T(\rho) = \frac{E \rho E}{Tr[\rho E]}, E \in \mathcal{P}(\mathcal{H})\}$. $(P^{\prime}(\mathbb{O}_{\mathbb{T}}), \leq, ^{\prime})$ is isomorphic to \\$( \mathcal{P}(\mathcal{H}),\leq, ^{\prime})$ as indicated by the following theorem. 
\vspace{-\baselineskip}
\begin{theorem}
If $(\mathcal{E}, \mathbb{S}, \mathbb{P},\mathbb{T}))$ is an event state operation structure then $(\mathbb{O}_{\mathbb{T}}, \circ, ^{\ast}, ^{\prime})$ is a Baer$ ^{\ast}$-Semigroup. The mapping $E \in \mathcal{E} \rightarrow T_{E} \in P(\mathbb{O}_{\mathbb{T}})$ is an isomorphism of the orthomodular orthoposet $(\mathcal{E},\leq, ^{\prime})$ onto the orthomodular orthoposet $(P^{\prime}(\mathbb{O}_{\mathbb{T}}),\leq, ^{\prime})$. 
\end{theorem}
\vspace{-\baselineskip}
\noindent For the proof we refer to \cite{pool1968baer}. 
\subsubsection{Compatibility and commutativity}\label{Algebraic structure of the set of events Event state operation structure Compatibility and commutativity}
\begin{theorem}
If $(\mathcal{E}, \mathbb{S}, \mathbb{P},\mathbb{T}))$ is an event state operation structure then $(\mathcal{E},\leq,^{\prime})$ is an ortholattice; furthermore, if $E_{1}, E_{2}  \in \mathcal{E}$ then \\ $T_{E_{1} \wedge E_{2}} =  (T_{E^{\prime}_{1}} \circ T_{E_{2}})^{\prime} \circ T_{E_{2}}$.
\end{theorem}
\vspace{-\baselineskip}
\noindent Proof: Since  $(P^{\prime}(\mathbb{O}_{\mathbb{T}}), \leq, ^{\prime})$ is an orthomodular ortholattice and $T_{E_{1}}$ and $T_{E_{2}} \in P^{\prime}(\mathbb{O}_{\mathbb{T}})$, from Theorem \ref{Theorem-1} it follows that $T_{E_{1}} \wedge T_{E_{2}} =  (T^{\prime}_{E_{1}} \circ T_{E_{2}}) ^{\prime} \circ T_{E_{2}}$. Since the mapping $E \in \mathcal{E} \rightarrow T_{E} \in P(\mathbb{O}_{\mathbb{T}})$ is an isomorphism of $(\mathcal{E},\leq, ^{\prime})$ onto $(P^{\prime}(\mathbb{O}_{\mathbb{T}}),\leq, ^{\prime})$, $T_{E_{1} \wedge E_{2}} = T_{E_{1}} \wedge T_{E_{2}}$ and $T_{E^{\prime}_{1}} = T^{\prime}_{E_{1}}$. Hence the result follows.
\vspace{-\baselineskip}
\begin{theorem}
If $(\mathcal{E}, \mathbb{S}, \mathbb{P},\mathbb{T}))$ is an event state operation structure and \\ $E_{1}, E_{2} \in \mathcal{E}$, then the following are equivalent:
\begin{inparaenum}
\item $E_{1} \mathcal{C} E_{2}$;\\
\item $T_{E_{1}}\circ T_{E_{2}} = T_{E_{2}}\circ T_{E_{1}}$.
\end{inparaenum}
If $E_{1} \mathcal{C} E_{2}$, then $T_{E_{1} \wedge E_{2}} = T_{E_{1}} \circ T_{E_{2}}$.
\end{theorem}
\vspace{-\baselineskip}
\noindent Proof: Let us define a new relation on the ortholattice $(P^{\prime}(\mathbb{O}_{\mathbb{T}}), \leq, ^{\prime})$ as $T \bar{\mathcal{C}} U$ if and only if $\exists \; T_{0}, U_{0}, V_{0} \in P^{\prime}(\mathbb{O}_{\mathbb{T}})$ such that $T_{0} \perp U_{0}$, $T_{0} \perp V_{0}$, $U_{0} \perp V_{0} $, $T = T_{0} \vee V_{0}$ and $U = U_{0} \vee V_{0}$. From Theorem \ref{Theorem-1}, it follows that $T \bar{\mathcal{C}} U$ if and only if $T \circ U = U \circ T$. Since the mapping $E \in \mathcal{E} \rightarrow T_{E} \in P(\mathbb{O}_{\mathbb{T}})$ is an isomorphism of $(\mathcal{E},\leq, ^{\prime})$ onto $(P^{\prime}(\mathbb{O}_{\mathbb{T}}),\leq, ^{\prime})$, $E_{1} \mathcal{C} E_{2}$ if and only if $T_{E_{1}} \bar{\mathcal{C}} T_{E_{2}}$. Hence $E_{1} \mathcal{C} E_{2}$ if and only if $T_{E_{1}}\circ T_{E_{2}} = T_{E_{2}}\circ T_{E_{1}}$. From Theorem \ref{Theorem-1} it  also follows that, if $E_{1} \mathcal{C} E_{2}$, then $T_{E_{1}}\circ T_{E_{2}} = T_{E_{2}}\circ T_{E_{1}}$, which implies that $T_{E_{1} \wedge E_{2}} = T_{E_{1}} \wedge T_{E_2}$.

\noindent Thus, we started of with a set of experimentally verifiable propositions whose elements we refer to as events. We were interested in understanding the algebraic structure of the set of events and then suitably construct a ``probability space" on it. We associated states, measures, and operations with the set of events. The set of events, implication relation on the set, and unary operation of orthocomplementation  on the set  was shown to be isomorphic to the set of closed operations with implication relation and unary operation. Hence the algebraic structure of the set of events is equivalent to the algebraic structure of the closed set of operations \cite{baras2003multiagent,  baras2016multiagent}. In the following problem, we infer the algebraic structure of the set of events by finding the algebraic structure of the set of operations. 
\section{Example: multi-agent decision making}\label{Example: multi-agent decision making}
\subsection{Problem description}\label{Example: multi-agent decision making Problem description}
We consider the binary hypothesis testing problem with three observers and a central coordinator. There are two possible states of nature. Each observer collects observations which are statistically related to the true state of nature. Following are the assumptions:
\begin{inparaenum}
\item The state of nature is the same for the three observers and the central coordinator;
\item Each observer knows the marginal distribution of the observations it alone collects; 
\item The joint distribution of the observations is unknown;
\item There is \textit{no} common notion of time for the observers. Each observer has a local notion of time; equivalently number of samples.  
\end{inparaenum}
Each observer constructs its own classical probability space (as discussed in \cite{raghavan2019binary}). The observers then formulate a sequential hypothesis testing problem in their respective probability spaces.  The sequential hypothesis testing problem is solved using SPRT. Let the decision of Observer $i$ be $D_{i}$.  The observers transmit their decision to the central coordinator. The decisions are received by the central coordinator. It is possible that the central coordinator  receives decisions from multiple observers simultaneously. \textit{We consider the scenario where the observer can collect (measure) only one observation at a given instant.} When multiple observations from different observers arrive simultaneously, then observations are collected with the following order of preference: Observer 2, followed by Observer 1 and then Observer 3. For e.g., if $D_{1}$ and $D_{2}$ arrive simultaneously,  then the observer measures $D_{2}$ first and then $D_{1}$. If all the three observations arrive simultaneously then $D_{2}$ is collected first followed by $D_{1}$ and then $D_{3}$. The objective of the central coordinator is to find its belief about the true state of nature by treating the decision information that it receives as observations. The central coordinator has to construct a suitable probability space where the hypothesis testing problems can be formulated and solved.   

\noindent Under either state of nature, the set of atomic propositions that can be verified by the central coordinator is $B = \{$ `$D_{1}$ is equal to 1', `$D_{1}$ is equal to 0', `$D_{2}$ is equal to 1', `$D_{2}$ is equal to 0', `$D_{3}$ is equal to 1', `$D_{3}$ is equal to 0', \textbf{0}, \textbf{1} $\}$. The propositions \textit{do not include} the time at which the decision was received.  We will elaborate more on this statement at the end of this section. Let $\bar{B}$ denote the set of experimentally verifiable events for the central coordinator. Clearly $B \subseteq \bar{B}$. At this juncture, $\bar{B}$ does not include the conjunction and the disjunction of the events in $B$. As discussed in the following sections, if some of the events are compatible then their conjunction and disjunction will be included as separate events in $\bar{B}$. 
 
\noindent \textit{Hypothesis: We hypothesize that the set of events along with the set inclusion as the relation of implication form a Boolean algebra and thus the states correspond to classical probability measures}. 

\noindent From our hypothesis it follows that $(\bar{B},\leq)$, where `$\leq$' is the set inclusion, is a Boolean algebra. $\bar{B}$ includes events of the form $E_{1} \wedge E_{2}$, $E_{1} \wedge E_{2} \vee E_{3}$, etc., and the distributive identity is satisfied. Assuming that the hypothesis is true, the central coordinator can construct an event state structure along the lines of \textit{Example 2}. The operation corresponding to an event $E \in \bar{B}$ (as in \textit{Example 2}) is defined as
\begin{align*}
(T_{E}(\rho))(F) = \frac{\rho( E \wedge F)}{\rho(E)}
\end{align*}
Since we are hypothesizing that the set of events form a Boolean lattice, it is expected that $T_{E} \circ T_{F}(\rho) = T_{F} \circ T_{E}(\rho)$ for all $E,F \in \bar{B}$ and for all states in the domain. We are interested in verifying if the event $E_{1}$ =`$D_{1}$ is equal to 1' and the event $E_{2}$=`$D_{2}$ is equal to 1' are compatible. Verifying $E_{1} \mathcal{C} E_{2}$ is equivalent to verifying $T_{E_{1}} \circ T_{E_{2}}(\rho) = T_{E_{2}}\circ T_{E_{1}} (\rho)$, for all states in the domain. Let $E_{3}$ = `$D_{3}$ is equal to 1'. If the equality of the two operations is indeed true, then $\mathbb{P}(T_{E_{1}} \circ T_{E_{2}}(\rho),E_{3}) = \mathbb{P}(T_{E_{2}} \circ T_{E_{1}}(\rho),E_{3})$.

\noindent At the central coordinator, there does not exist a stochastic model (as in Example 2 or Example 3 ). Hence, the initial state $\rho$ (for the classical model or the noncommutative model) is unknown and so are the the projections corresponding to events $E_1$, $E_2$ or $E_3$ (in the noncommutative model). We consider a learning phase for the central coordinator. During the learning phase of the central coordinator the true state of nature is known only to the central coordinator. The three observers perform sequential hypothesis test individually and transmit their beliefs to the central coordinator.  For every experiment, the central coordinator collects a vector of the form $(h,d_1,d_2,d_3)$, where $h$ is the true state of nature and $d_1,d_2,d_3$ are decisions from the three observers in the order they are collected. 

\begin{definition}
Empirical distribution is a set function, $ \mathbb{G}: \mathcal{E} \rightarrow [0,1]$. Given that an experiment has been repeated $N$ times, for an event $E \in \mathcal{E}$, 
\begin{align*}
\mathbb{G}(E) = \frac{\text{Number of occurrences of E}}{N}
\end{align*}
\end{definition}
\noindent Using the above definition, $\mathbb{P}(T_{E_{1}} \circ T_{E_{2}}(\rho),E_{3})$ is estimated as 
\begin{align*}
&\mathbb{G}(E_{3} \; \text{given} \; E_{2} \; \text{as occurred first followed by} E_{1}) \\ = &\frac{\text{No. of occurrences of $D_{2}=1$ followed by $D_{1} =1$ and then $D_{3}=1$}}{\text{No. of occurrences of $D_{2}=1$ followed by $D_{1} =1$}}
\end{align*}
\noindent It is assumed that for $N$ sufficiently large, $\mathbb{P}(T_{E_{1}} \circ T_{E_{2}}(\rho),E_{3})$ will be arbitrarily close to the empirical value calculated as above. 

\noindent We consider a specific setup for the above problem. The marginal distributions learned by the three observers (under either state of nature) have been listed in Table \ref{Table 1}. Simulations were performed to develop a suitable stochastic model for the central coordinator. In each simulation run, the three observers receive observations drawn from their respective distributions. Based on the observations and sequential hypothesis test, they arrive at a belief about the true state of nature which they send to the central coordinator. $10^{7}$ such simulations were performed.  
\begin{table}[h!]
\begin{center}
\begin{tabular}{||c || c|| c ||} 
\hline
$X$ & $h=0$ & $h=1$\\
\hline
$1$ & $0.20$ & $0.40$\\
\hline
$2$ & $0.10$ & $0.20$\\
\hline
$3$ & $0.15$ & $0.10$\\
\hline
$4$ & $0.30$ & $0.15$\\
\hline
$5$ & $0.25$ & $0.15$\\
\hline
\end{tabular}
\quad
\begin{tabular}{||c || c|| c ||} 
\hline
$Y$ & $ h=0$ & $ h=1$\\
\hline
$1$ & $0.20$ & $0.25$\\
\hline
$2$ & $0.40$ & $0.30$\\
\hline
$3$ & $0.30$ & $0.20$\\
\hline
$4$ & $0.10$ & $0.25$\\
\hline
\end{tabular}
\quad 
\begin{tabular}{||c || c|| c ||} 
\hline
$Z$ & $ h=0$ & $ h=1$\\
\hline
$1$ & $0.25$ & $0.35$\\
\hline
$2$ & $0.35$ & $0.50$\\
\hline
$3$ & $0.40$ & $0.15$\\
\hline
\end{tabular}
\vspace{0.2cm}
\caption{Distribution of observations under either hypothesis for Observer 1, Observer 2 and Observer 3}	
\label{Table 1}
\end{center}
\vspace{-\baselineskip}
\end{table}  
$T_{E_{2}} \circ T_{E_{1}} (\rho)$ is the state conditioned on the event $E_{1}$ and then the event $E_{2}$ and the state $\rho$. As discussed previously, $\mathbb{P}(T_{E_{2}} \circ T_{E_{1}}(\rho),E_{3})$ is estimated as follows. Let $\alpha$ be the number of simulations in which $D_{1} =1$, followed by $D_{2} =1$, and then $D_{3} = 1$.  Let $\beta$ be the number of simulations in which $D_{1} =1$, followed by $D_{2} =1 $, and then $D_{3} = 0$. Then $\mathbb{P}(T_{E_{2}} \circ T_{E_{1}}(\rho),E_{3}) = \frac{\alpha}{\alpha + \beta}$.The probabilities in Tables \ref{Table 2} and \ref{Table 3} have been estimated using the relative frequency approach. 

\noindent From Tables \ref{Table 2} and \ref{Table 3}, we infer that under either state of nature, for some state $\rho$ in the domain, $T_{E_{2}} \circ T_{E_{1}} (\rho) \neq T_{E_{1}} \circ T_{E_{2}}(\rho) $. Hence events $E_{1}$ and $E_{2}$ are not compatible. For the same marginal distributions for the observers, it was observed that pairs $E_{1}, E_{3}$ and $E_{2}, E_{3}$ were incompatible. The set of experimentally verifiable events $\bar{B}$ is equal to $B$.  Our initial hypothesis that the set of events form a Boolean algebra is incorrect. Instead, the set of events form an orthomodular ortholattice as discussed in the next section. 
\begin{table}[ht!]
\vspace{-\baselineskip}
\begin{center}
\begin{tabular}{||c || c|| c ||} 
\hline
$\mathbb{P}(T_{(\cdot)} \circ T_{(\cdot)}(\rho),(\cdot))$& $E^{\prime}_{3}$ & $E_{3}$\\
\hline
$T_{E^{\prime}_{1}} \circ T_{E^{\prime}_{2}}(\rho)(\cdot)$& $0.5880$ & $0.4120$\\
\hline
$T_{E^{\prime}_{1}} \circ T_{E_{2}}(\rho)(\cdot)$& $0.5203$ & $0.4797$\\
\hline
$T_{E_{1}} \circ T_{E^{\prime}_{2}}(\rho)(\cdot)$& $0.5915$ & $0.4085$\\
\hline
$T_{E_{1}} \circ T_{E_{2}}(\rho)(\cdot)$& $0.5372$ & $0.4628$\\
\hline
\end{tabular}
\begin{tabular}{||c || c|| c ||} 
\hline
$\mathbb{P}(T_{(\cdot)} \circ T_{(\cdot)}(\rho),(\cdot))$& $E^{\prime}_{3}$ & $E_{3}$\\
\hline
$T_{E^{\prime}_{2}} \circ T_{E^{\prime}_{1}}(\rho)(\cdot)$& $0.6113$ & $0.3887$\\
\hline
$T_{E_{2}} \circ T_{E^{\prime}_{1}}(\rho)(\cdot)$& $0.6026$ & $0.3974$\\
\hline
$T_{E^{\prime}_{2}} \circ T_{E_{1}}(\rho)(\cdot)$& $0.6154$ & $0.3846$\\
\hline
$T_{E_{2}} \circ T_{E_{1}}(\rho)(\cdot)$& $0.6095$ & $0.3905$\\
\hline
\end{tabular}
\vspace{0.2cm}
\caption{Conditional probabilities when true hypothesis is zero}	
\label{Table 2}
\end{center}
\vspace{-\baselineskip}
\end{table} 
\vspace{-\baselineskip}
\begin{table}[h!]
\vspace{-\baselineskip}
\begin{center}
\begin{tabular}{||c || c|| c ||} 
\hline
$\mathbb{P}(T_{(\cdot)} \circ T_{(\cdot)}(\rho),(\cdot))$& $E^{\prime}_{3}$ & $E_{3}$\\
\hline
$T_{E^{\prime}_{1}} \circ T_{E^{\prime}_{2}}(\rho)(\cdot)$& $0.1547$ & $0.8453$\\
\hline
$T_{E^{\prime}_{1}} \circ T_{E_{2}}(\rho)(\cdot)$& $0.1184$ & $0.8816$\\
\hline
$T_{E_{1}} \circ T_{E^{\prime}_{2}}(\rho)(\cdot)$& $0.1530$ & $0.8470$\\
\hline
$T_{E_{1}} \circ T_{E_{2}}(\rho)(\cdot)$& $0.1220$ & $0.8780$\\
\hline
\end{tabular}
\begin{tabular}{||c || c|| c ||} 
\hline
$\mathbb{P}(T_{(\cdot)} \circ T_{(\cdot)}(\rho),(\cdot))$& $E^{\prime}_{3}$ & $E_{3}$\\
\hline
$T_{E^{\prime}_{2}} \circ T_{E^{\prime}_{1}}(\rho)(\cdot)$& $0.1569$ & $0.8431$\\
\hline
$T_{E_{2}} \circ T_{E^{\prime}_{1}}(\rho)(\cdot)$& $0.1518$ & $0.8482$\\
\hline
$T_{E^{\prime}_{2}} \circ T_{E_{1}}(\rho)(\cdot)$& $0.1595$ & $0.8405$\\
\hline
$T_{E_{2}} \circ T_{E_{1}}(\rho)(\cdot)$& $0.1560$ & $0.8440$\\
\hline
\end{tabular}
\vspace{0.2cm}
\caption{Conditional probabilities when true hypothesis is one}	
\label{Table 3}
\end{center}
\vspace{-\baselineskip}
\vspace{-\baselineskip}
\end{table}
\vspace{0cm}
\subsection{Probability space construction}\label{Example: multi-agent decision making Probability space construction}

Let us consider the construction of the von Neumann Hilbert space model for the central co-ordinator. Let $\mathcal{H} = \mathbb{R}^{2}$. Let $\mathcal{P}(\mathbb{R}^{2})$ denote the set of orthogonal projections onto $\mathcal{H}$. Let $\mathbb{S}$ denote the set of symmetric, positive semidefinite matrices whose trace is 1. Let $E_{i}$, $i=1,2,3 \in \mathcal{P}(\mathbb{R}^{2})$ denote the projections of rank one corresponding to the events `$D_{i}$ is equal to 1'. The projections do not commute, $E_{i}E_{j} \neq E_{j}E_{i}$. The set of events  is $\mathcal{E} = \{E_{1}, E_{2}, E_{3}, \\ I_{\mathbb{R}^{2}}-E_{1}, I_{\mathbb{R}^{2}} - E_{2},  I_{\mathbb{R}^{2}} - E_{3}, \Theta, I_{_{\mathbb{R}^{2}}}\}$. The probability function is defined as $\mathbb{P}(\rho, E)  = Tr[\rho E] $. The event-state structure constructed for the central coordinator corresponds to $(\mathcal{E}, \mathbb{S}, \mathbb{P})$. The definition of the relation of implication is retained, i.e., $E_{1} \leq E_{2}$ if and only if $\mathbb{S}_{1}(E_{1}) \subseteq \mathbb{S}_{1}(E_{2})$. It is clear that $\mathcal{E}$ is a lattice as $ E\wedge F = \theta, \; \text{and} \; E\vee F =I, E, F \in \mathcal{E}, E \neq F$. Let $\mathbb{O}$ denote the set of all mappings from $\mathbb{S}$ to $\mathbb{S}$. The operation conditioned on an event is defined as $T_{E} = \frac{E\rho E}{Tr[\rho E]}$, as defined in Example 3.  Let $\mathbb{O}_{\mathbb{T}}$  be the set of operations of the form $T = T_{F_{1}} \circ T_{F_{2}} \circ \ldots \circ T_{F_{n}}, F_{1}, F_{2}, \ldots, F_{n} \in \mathcal{E}$. For an operation $T= T_{F_{1}} \circ T_{F_{2}} \circ \ldots \circ T_{F_{n}}$, the event corresponding to the orthocomplementation is the projection on to nullspace of $F_{n}F_{n-1}\ldots F_{1}$. If for some $i$, $F_{i}F_{i+1}$ is such that $F_{i+1} = I_{\mathbb{R}^{2}}- F_{i}$, then $F_{n}F_{n-1}\ldots F_{1} = \Theta$.  In such a case the projection is $I_{\mathbb{R}^{2}}$. Else, $\mathcal{R}(F_{1})$ is not orthogonal to $\mathcal{R}(F_{2})$. $F_{2}F_{1}(h) \neq \theta$ for \\ $h \notin \mathcal{R}(I-F_{1})$. Suppose for $F_{m}F_{m-1}\ldots F_{1}(h)\neq \theta$ for $h \notin \mathcal{R}(I-F_{1})$ for \\ some $m$, $2 \leq m \leq n-1$. $F_{m}F_{m-1}\ldots F_{1}(h) \in \mathcal{R}(F_{m})$. Since $\mathcal{R}(F_{m})$ is not orthogonal to $\mathcal{R}(F_{m+1})$, $F_{m+1}F_{m}\ldots F_{1}(h)\neq \theta$. Thus, $F_{n}F_{n-1}\ldots F_{1}(h)\neq \theta$ for  $h \notin \mathcal{R}(I-F_{1})$. $F_{n}F_{n-1}\ldots F_{1}(h) = \theta$ for $h \in \mathcal{R}(I-F_{1})$. Hence the projection is $I-F_{1}$. The other axioms associated with operations can be verified. The set of operations, the composition of operations, involution, and orthocomplmentation, $(\mathbb{O}_{\mathbb{T}}, \circ, ^{\ast}, ^{\prime})$, form a noncommutative Baer$^{\ast}$ semigroup. The set of closed projections, composition, and orthocomplememtation, \\ $(P^{\prime}(\mathbb{O}_{\mathbb{T}}), \leq, ^{\prime})$ is an orthomodular ortholattice. Since $(P^{\prime}(\mathbb{O}_{\mathbb{T}}), \leq, ^{\prime})$ is isomorphic to $(\mathcal{E}, \leq, ^{\prime})$, $(\mathcal{E}, \leq, ^{\prime})$ is an orthomodular ortholattice.

\subsection{Discussion}\label{Example: multi-agent decision making Discussion }

Suppose the three observers and the central coordinator have a common notion of time and the joint distribution of the measurements collected by the three observers is known. We can then construct a common probability space for the three agents and the central coordinator. Let $\tau_{i}$ denote the stopping time of Observer $i$. $\tau_{1}$, $\tau_{2}$, and $\tau_{3}$ are random variables in the common probability space. Let $D_{i}$ denote the decision of observer $i$ at stopping time $\tau_{i}$. Suppose the central coordinator can collect multiple  observations simultaneously, i.e., when $\tau_{1} = \tau_{2} = \tau_{3}$ (or $\tau_{i} = \tau_{j}, i \neq j$) then the central coordinator can simultaneously collect $D_{1}, D_{2}$ and $D_{3}$ (or $D_{i} \; \text{and} \; D_{j}$). In this scenario, when the joint distribution is known and the central coordinator is able to simultaneously collect observations from different observers, the concern of order effects does not arise. Different orders of measurement correspond to specific events in the sigma algebra. When the true state of nature is 1, $\mathbb{P}(T_{E_{2}} \circ T_{E_{1}}(\rho),E_{3})=\mathbb{P}(D_{3} = 1 | D_{2} = 1,D_{1} = 1, \tau_{3} \geq \tau_{2} > \tau_{1}, H=1)$ and $\mathbb{P}(T_{E_{1}} \circ T_{E_{2}}(\rho),E_{3})=\mathbb{P}(D_{3} = 1 | D_{2} = 1,D_{1} = 1, \tau_{3} \geq \tau_{1} \geq \tau_{2}, H=1)$. It is \\ not necessary that $\mathbb{P}(D_{3} = 1 | D_{2} = 1,D_{1} = 1, \tau_{3} \geq \tau_{2} > \tau_{1}, H=1)$ equals \\ $\mathbb{P}(D_{3} = 1 | D_{2} = 1,D_{1} = 1, \tau_{3} \geq \tau_{1} \geq \tau_{2}, H=1)$. In the absence of the joint distribution, when the probabilities $\mathbb{P}(T_{E_{2}} \circ T_{E_{1}}(\rho),E_{3})$ and \\ $\mathbb{P}(T_{E_{1}} \circ T_{E_{2}}(\rho), E_{3})$ are estimated from samples one could expect the ``order effects" to occur. The information (or knowledge) available to the central coordinator, its inability /ability to collect different observations simultaneously and the asynchrony in the observations plays an important role in determining the presence or absence of order effects. In our previous work \cite{raghavan2019binary, raghavan2019thesis}, we considered two synchronous observers with specific observation and information exchange pattern. Each observer either collects an observation or receives information from the other agent, but not both. Hence the issue of ``simultaneous verifiability" does not arise and the order effect was not observed. In the situation where the joint distribution is not available but the central coordinator is able to collect multiple observations simultaneously, it might be possible to construct a classical probability space by considering events of the form `time = k and $D_{1} =1$', `time = k and $D_{1}$ is unknown ', etc. This case requires further investigation.    
    
\noindent Our original goal was to study the hypothesis testing problem at the central coordinator.  Given the noncommutative probability space, we now discuss how hypothesis testing problems can be formulated and solved in such spaces. 
\section{Binary hypothesis testing problem}\label{Binary hypothesis testing problem}
\subsection{Problem formulation}\label{Binary hypothesis testing problem Problem formulation}
We consider a single observer. The observation collected by the observer is denoted by $Y$, $Y \in S$, $|S| = N$ where $S$ is a finite set of real numbers or real vectors of finite dimension. A fixed number of data strings consisting of observation and true hypothesis are collected by the observer. From the data strings, empirical distributions are found. Let $p^{h}_{i}, 1 \leq i \leq N$ be the distribution under hypothesis $h$. The prior probabilities of hypotheses can be found from the data and are represented by $\zeta_{1}$ (for $H=1$) and $\zeta_{0}$ \\ (for $H=0$). In the quantum probability framework, there are multiple ways in which measurements can be captured. Two of them are: (a) Projection valued measures  (PVM); (b) Positive operator valued measures (POVM). In this section we discuss the formulation of the detection problem in the classical probability framework and in then von Neumann probability framework with both representations for measurements.
\subsubsection{Classical probability}\label{Binary hypothesis testing problem Problem formulation Classical probability}
Let $\Omega = \{0,1\} \times S$ be the sample space. Let $\mathcal{F} = 2^{\Omega}$ be the associated $\sigma$ algebra. An element in the sample space can be represented by $\omega=(h,y)$, where $h\in \{0,1\}$ and $y \in  S$. The measure is $\mathbb{P}(\omega) = \zeta_{h}p^{h}_{y}$. The probability space is $(\Omega,\mathcal{F},\mathbb{P})$. Given a new observation, $Y=y$ the detection problem is to find $D$ such that the following cost is minimized:
\begin{align*}
\mathbb{E}_{\mathbb{P}} [H(1-D)+ (1-H)D],
\end{align*}
i.e, the probability of error is minimized. $H$ represents the hypothesis random variable. Once the decision is found the optimal cost also needs to be found. 
\subsubsection{Projection valued measure}\label{Binary hypothesis testing problem Problem formulation Projection valued measure}
\begin{definition}
Projection Valued Measure(PVM): Let $(X, \Sigma)$ be a measurable space. A projection valued measure is a mapping $F$ from $\Sigma$ on to $\mathcal{P}(\mathcal{H})$ such that,
\begin{inparaitem}
\item[(i)] $F(X) = \mathbb{I}$;
\item[(ii)] $A, B \in \Sigma$ such that $ A \cap B =  \emptyset$, then $F(A \cup B) =  F(A) + F(B)$;
\item[(iii)] If $\{A_{i}\}_{i \geq 1} \subseteq \Sigma$, such that $A_{1} \subset A_{2} \subset ...$, then $F(\cup^{\infty}_{i=1}A_{i}) =  \underset{i \rightarrow \infty} \lim F(A_{i})$.
\end{inparaitem}
\end{definition}
\vspace{-\baselineskip}
\noindent  For the detection problem, $X = \{1,2,...,N\}$, $\Sigma  = 2^{X}$. The second condition implies that the minimum dimension of the complex Hilbert space in consideration is $N$. We let $\mathcal{H} = \mathbb{C}^{N}$. The first objective is to find $\rho_{h} \in \mathcal{T}^{+}_{s}(\mathbb{C}^{N})$, $h =0,1$ and $F:\Sigma \rightarrow \mathcal{P}(\mathbb{C}^{N})$,  such that:
\begin{inparaitem}
\item[(i)] $Tr[\rho_{h}F(i)] = p^{h}_{i}, h= 0, 1, 1 \leq i \leq N,\; \label{Equation 1}$ ;
\item[(ii)] $F(i)F(j) = \Theta_{\mathbb{C}^{N}}\;, 1 \leq i, j \leq N, i \neq j \; \text{and} \; \sum^{N}_{i=1}F(i) =  \mathbb{I}_{\mathbb{C}^{N}}, \label{Equation 2}$,
\end{inparaitem}
where $\Theta_{\mathbb{C}^{N}}$ is the zero operator and $\mathbb{I}_{\mathbb{C}^{N}}$ is the identity operator. Given the state and the PVM, we consider the formulation of the detection problem mentioned in  \cite{baras1979noncommutative}, section 3.4. Let $C_{ij}$ denote the cost incurred when the decision made is $i$ while the true hypothesis is $j$. Since the objective is to minimize the probability of error, we let $C_{10} = 1$, $C_{00} = 0$, $C_{01} = 1$ and $C_{11} = 0$. The decision policy, $\{\alpha^{1}_{i} \}^{N}_{i=1}$ and $\{\alpha^{0}_{i} \}^{N}_{i=1}$ denotes the probability of choosing $D=1$  and $D=0$ respectively when observation $i$ is received. Given observation $i$, the probability of choosing $D=1$ ($D$ is the decision) and the true hypothesis being $0$ is $\zeta_{0}Tr[\rho_{0}F(i)]\alpha^{1}_{i}$. Hence, the probability of choosing $D=1$ and true hypothesis being $0$ is $\sum^{N}_{i=1}\zeta_{0}Tr[\rho_{0}F(i)]\alpha^{1}_{i} $. Similarly, the probability of choosing $D=0$ and true hypothesis being $1$ is $\sum^{N}_{i=1}\zeta_{1}Tr[\rho_{1}F(i)]\alpha^{0}_{i} $. The probability of error is:
\begin{align*}
\mathbb{P}_{e} = Tr\left[\zeta_{0}\rho_{0}\left[\sum^{N}_{i=1}\alpha^{1}_{i} F(i)\right] + \zeta_{1}\rho_{1}\left[\sum^{N}_{i=1}\alpha^{0}_{i} F(i)\right]\right].
\end{align*}
We define the risk operators and note that,
\begin{align*}
W_{1} = \zeta_{0}\rho_{0},\; W_{0} = \zeta_{1}\rho_{1}, \sum^{N}_{i=1}\alpha^{h}_{i} F(i) \geq 0, h = 0,1\sum^{N}_{i=1}\alpha^{1}_{i}F(i) +\alpha^{0}_{i} F(i) = \mathbb{I}_{\mathbb{C}^{N}}.
\end{align*}
Instead of minimizing over the decision policies, we minimize over pairs of operators which are semi-definite and sum to identity. Hence, the detection problem is formulated as follows 
\begin{align*}
P1 : \underset{\Pi_{1}, \Pi_{0}} \min Tr[W_{1}\Pi_{1} + W_{0}\Pi_{0}] \; \text{s.t} \; \Pi_{1}, \Pi_{0} \in \mathcal{B}^{+}_{s}(\mathbb{C}^{N}), \Pi_{1} + \Pi_{0} = \mathbb{I}_{\mathbb{C}^{N}}.
\end{align*}
The solution of the above problem, $\Pi^{*}_{1}$, $\Pi^{*}_{0}$ are the detection operators which are to be realized using the PVM:
\begin{align*}
P2: \exists \{ \alpha^{1}_{i} \}^{N}_{i=1} \; \text{s.t} \; 0 \leq \alpha^{1}_{i} \leq 1, \; \forall i, \Pi^{*}_{1}= \sum^{n}_{i=1} \alpha^{1}_{i}F(i), \Pi^{*}_{0}= \sum^{n}_{i=1} (1-\alpha^{1}_{i})F(i).
\end{align*}
Suppose for two pairs of states, $(\rho_{1},\rho_{0})$, $(\bar \rho_{1},\bar \rho_{0})$ and PVM $F$, satisfy the condition below, 
\begin{align*}
Tr[\rho_{h}F(i)] = Tr[\bar \rho_{h}F(i)]  =  p^{h}_{i}, h= 0, 1, 1 \leq i \leq N. 
\end{align*}
If we consider the solution to P1 alone, the corresponding detection operators $(\Pi^{*}_{1}$, $\Pi^{*}_{0})$, $(\bar \Pi^{*}_{1}$, $\bar \Pi^{*}_{0})$ and the respective minimum costs achieved, $\mathbb{P}_{e}, \bar{\mathbb{P}}_{e}$ could be different. However, if we consider solution to P1 such that P2 is feasible, i.e., the detection operators are realizable, then, \\ $\mathbb{P}_{e} =\sum^{N}_{i=1}\zeta_{0}Tr[\rho_{0}F(i)]\alpha^{1}_{i} + \sum^{N}_{i=1}\zeta_{1}Tr[\rho_{1}F(i)]\alpha^{0}_{i}= \\ \sum^{N}_{i=1}\zeta_{0}Tr[\bar \rho_{0}F(i)]\alpha^{1}_{i} + \sum^{N}_{i=1}\zeta_{1}Tr[\bar \rho_{1}F(i)]\alpha^{0}_{i} \geq \bar{\mathbb{P}}_{e}$. \\ Similarly, $\bar{\mathbb{P}}_{e} \geq \mathbb{P}_{e}$. Hence $\bar{\mathbb{P}}_{e} = \mathbb{P}_{e}$. For a given PVM, the optimal cost does not change with different states that achieve the empirical distribution. 
\subsubsection{Positive operator valued measure}\label{Binary hypothesis testing problem Problem formulation Positive operator valued measure}
Consider the scenario the observer collects two observations, $Y = [Y_{1},Y_{2}]$. Let $Y_{1} \in Z_{1}$, $|Z_{1}| = \eta_{1}$ and $Y_{2} \in Z_{2}$, $|Z_{2}| = \eta_{2}$. Then $Y_{1}$ and $Y_{2}$ can be individually represented as PVMs in a Hilbert space of dimension $\eta$, $\eta = \max\{\eta_{1}, \eta_{2}\}$. Let the PVM corresponding to $Y_{1}$ and $Y_{2}$ be $\mu$ and $\nu$ respectively. Let the state be $\rho$. Suppose $Y_{1}$ is measured first and the value obtained is $i \in Z_{1}$. Then the state after measurement of $Y_{1}$ changes from $\rho$ to (\cite{davies1976quantum}):
\begin{align*}
\rho_{i} = \frac{\mu(i)\rho\mu(i)}{Tr[\rho \mu(i)]}. 
\vspace{0.2cm}
\end{align*} 
After measuring $Y_{1}$, $Y_{2}$ is measured. The conditional probability of $Y_{2} =j$ given $Y_{1} = i$ is,
\begin{align*}
Tr[\rho_{i}\nu(j)] = \frac{Tr[\mu(i)\rho\mu(i)\nu(j)]}{Tr[\rho \mu(i)]} = \frac{Tr[\rho\mu(i)\nu(j)\mu(i)]}{Tr[\rho \mu(i)]}. 
\end{align*}
The probability of obtaining $Y_{1} =i$ and then $Y_{2} = j$ is $Tr[\rho\mu(i)\nu(j)\mu(i)]$.
Further, the measurement corresponding to $Y$ is, $\sigma_{1} (i,j) = \mu(i)\nu(j)\mu(i),\\ 1 \leq i \leq \eta_{1}, 1\leq j \leq \eta_{2}$.  If $Y_{1}$ is measured after $Y_{2}$, then the measurement corresponding to $Y$ is, $\sigma_{2} (i,j) = \nu(i)\mu(j)\nu(i), 1 \leq i \leq \eta_{2}, 1\leq j \leq \eta_{1}$. If for any $(i,j)$,  $\mu(i)$ and $\nu(j)$ do not commute, $\sigma_{1} (i,j)$ and $\sigma_{2} (i,j)$ are not projections. They are positive, Hermitian and bounded. Hence $\sigma_{1}$, $\sigma_{2}$ are not PVMs,  and belong to a larger class of measurements, i.e., they are  POVMs.
\vspace{-\baselineskip}
\begin{definition}
Positive Operator Valued Measure (POVM): Let $(X, \Sigma)$ be a measurable space. A positive operator valued measure is a mapping $M$ from $\Sigma$ on to $\mathcal{B}^{+}_{s}(\mathcal{H})$ such that, if $\{X_{i}\}_{i \geq 1}$ is partition of $X$, then  
\begin{align*}
\sum_{i} M(X_{i}) = \mathbb{I} \;\;(\text{Strong Operator Topology})
\end{align*}
\end{definition}
\vspace{-\baselineskip}
\noindent Further, for $A, B \in \Sigma$ such that $ A \cap B =  \emptyset$, if $M(A)M(B) = \Theta_{\mathcal{H}}$, then $M$ is a PVM. 

\noindent We consider the dimension of the Hilbert space to be $k$, $k \geq 2$.  As in the previous formulation, the first objective is to find states, $\hat{\rho}_{h} \in \mathcal{T}^{+}_{s}(\mathbb{C}^{k}) ,\\ h=0,1$ and POVM, $M: \Sigma \rightarrow \mathcal{B}^{+}_{s}(\mathbb{C}^{k})$ such that
\begin{align}
\hspace{-0.2cm}Tr[\hat{\rho}_{h}M(i)] = p^{h}_{i}, h= 0, 1, 1 \leq i \leq N \;  \text{and} \sum^{N}_{i=1} M(i) = \mathbb{I}_{\mathbb{C}^{k}}.\label{Equation 3}
\end{align} 
The probability of error calculation is analogous to the previous section. We define the new risk operators as $\hat W_{1} = \zeta_{0}\hat{\rho}_{0},\; \hat W_{0} = \zeta_{1}\hat \rho_{1}$. Given states and POVM, the detection problem with the same cost parameters as P1, is formulated as: 
\begin{align*}
P3: \underset{\hat \Pi_{1}, \hat \Pi_{0}} \min Tr[\hat W_{1} \hat \Pi_{1} + \hat W_{0} \hat \Pi_{0}] 
\; \text{s.t} \; \hat \Pi_{1}, \hat \Pi_{0} \in \mathcal{B}^{+}_{s}(\mathbb{C}^{k}), \; \hat \Pi_{1} + \hat \Pi_{0} = \mathbb{I}_{\mathbb{C}^{k}}.
\end{align*}
The decision policies $\{ \beta^{1}_{i} \}^{N}_{i=1}$ and $\{ \beta^{0}_{i} \}^{N}_{i=1}$ are found by solving the following problem:
\begin{align*}
P4: \exists \{ \beta^{1}_{i} \}^{N}_{i=1} \;\text{s.t}\; 0 \leq \beta^{1}_{i} \leq 1 , \hat \Pi_{1}= \sum^{n}_{i=1} \beta^{1}_{i}M(i), \; \hat \Pi_{0}= \sum^{n}_{i=1} (1-\beta^{1}_{i})M(i). 
\end{align*}
Consider the problem:
\begin{align*}
&P5: \underset{\hat \Pi_{1}, \hat \Pi_{0},\{ \beta^{1}_{i} \}^{N}_{i=1} } \min  Tr[\hat W_{1} \hat \Pi_{1} + \hat W_{0} \hat \Pi_{0}] \text{s.t} \hat \Pi_{1},\hat \Pi_{0} \in \mathcal{B}^{+}_{s}(\mathbb{C}^{k}), \;\hat \Pi_{1} + \hat \Pi_{0} = \mathbb{I}_{\mathbb{C}^{k}}, \\
&0 \leq \beta^{1}_{i} \leq 1, 1 \leq i \leq n, \hat \Pi_{1}= \sum^{n}_{i=1} \beta^{1}_{i}M(i), \; \hat \Pi_{0}= \sum^{n}_{i=1}(1-\beta^{1}_{i})M(i). 
\end{align*}
Let the feasible set of detection operators for $P3$ be $S_{1}$ and for $P5$ be $S_{2}$. Due to additional constraints in $P5$, $S_{2} \subseteq S_{1}$. The detection operators obtained by solving $P3$ may or may not be realizable, i.e., $P4$ may not be feasible. In $P5$, the optimization is only over detection operators which are realizable.  If the solution of $P3$ is such that $P4$ is feasible then it is the solution for $P5$ as well. It is also possible that $P3$ is solvable, $P4$ is not feasible and $P5$ is solvable. The objective is to understand the minimum probability of error which can be achieved by detection operators which are realizable. Hence, we consider the solution of $P5$ and compare it with the minimum error achieved in the PVM formulation. 

\noindent Let $\mathbb{M}$ be set of all POVMs on $\Sigma$. Let $\hat{\mathbb{S}} \subset \mathcal{T}^{+}_{s}(\mathbb{C}^{k}) \times \mathcal{T}^{+}_{s}(\mathbb{C}^{k}) \times \mathbb{M}$ be the set of, pairs of states and a POVM such that equation (\ref{Equation 3}) is satisfied. Let $\bar{\mathbb{S}} \subseteq \hat{\mathbb{S}}$ be the triples for which the optimization problem $P5$ can be solved. For a triple $(\hat{\rho}_{0}, \hat{\rho}_{1},M)$ in $\bar{\mathbb{S}}$, we define $Q(\hat{\rho}_{0}, \hat{\rho}_{1},M)$ to be the optimal value achieved by solving $P5$.
\vspace{-\baselineskip}
\vspace{-\baselineskip}
\subsection{Solution}\label{Binary hypothesis testing problem Solution}
\subsubsection{Classical probability}\label{Binary hypothesis testing problem Solution Classical probability}
It suffices to minimize, \\ $\mathbb{E}_{\mathbb{P}} [H(1-D)+ (1-H)D |Y=y]$. We note the conditional expectation can be computed as $\mathbb{E}_{\mathbb{P}} [H|Y=y] = \frac{p^{1}_{y}\zeta_{1}}{p^{1}_{y}\zeta_{1}+p^{0}_{y}\zeta_{0}}$. The solution for the optimization problem is a threshold policy: $D=1$ if $p^{1}_{y}\zeta_{1} \geq p^{0}_{y}\zeta_{0}$ else $D= 0$.\\ Thus, the cost when the observation is $y$ is $\frac{\min\{p^{1}_{y}\zeta_{1},p^{0}_{y}\zeta_{0}\}}{p^{1}_{y}\zeta_{1}+p^{0}_{y}\zeta_{0}}$. The expected cost is:
\begin{align*}
\sum^{N}_{i=1}\left[\frac{\min\{p^{1}_{i}\zeta_{1},p^{0}_{i}\zeta_{0}\}}{p^{1}_{i}\zeta_{1}+p^{0}_{i}\zeta_{0}}\right] \times \mathbb{P}(Y = i) = \sum^{N}_{i=1}\min\{p^{1}_{i}\zeta_{1},p^{0}_{i}\zeta_{0}\}.
\end{align*}
\subsubsection{Projection valued measure}\label{Binary hypothesis testing problem Solution Projection valued measure}
Define, 
\[
  \rho_{h} =
  \begin{bmatrix}
    p^{h}_{1} & & \\
    & \ddots & \\
    & & p^{h}_{N}
  \end{bmatrix}
  \;\text{and}\; F(i)= e_{i}e_{i}^{H},
\]
where $e_{i}$ represents the canonical basis in $\mathbb{C}^{N}$. Clearly conditions mentioned in section  \ref{Equation 1} are satisfied. 

\begin{theorem}
[\cite{baras1979noncommutative, baras1987distributed, baras2016multiagent}] There exists a solution to the problem, minimize $Tr[W_{0}\Pi_{0} + W_{1}\Pi_{1}]$ over all two component POM's, where $W_{0}, W_{1}$ belong to  $\mathcal{B}^{+}_{s}(\mathbb{C}^{N})$. A necessary and sufficient condition for $\Pi^{*}_{i}$ to be optimal is that:
\begin{align*}
W_{0}\Pi^{*}_{0} + W_{1}\Pi^{*}_{1} \leq W_{i}, i =0, 1, \; \Pi^{*}_{0}W_{0} + \Pi^{*}_{1}W_{1} \leq W_{i}, i =0, 1 
\end{align*}
Furthermore, under any of above conditions the operator 
\begin{align*}
 O = W_{0}\Pi^{*}_{0} + W_{1}\Pi^{*}_{1} \\ = \Pi^{*}_{0}W_{0} + \Pi^{*}_{1}W_{1},   
\end{align*}
is self-adjoint and the unique solution to the dual problem.
\end{theorem}
\noindent To solve P1, we invoke the above theorem. $\Pi^{*}_{1}$ and $\Pi^{*}_{0}$ solve $P1$ and $P2$ can be solved if they satisfy the following conditions: $W_{1}\Pi^{*}_{1} + W_{0}\Pi^{*}_{0}   \leq W_{1},\\ W_{1}\Pi^{*}_{1} + W_{0}\Pi^{*}_{0}  \leq W_{0}, \Pi^{*}_{1}, \Pi^{*}_{0} \in \; \mathcal{B}^{+}(\mathbb{C}^{N}), \Pi^{*}_{1} + \Pi^{*}_{0}=  \mathbb{I}_{\mathbb{C}^{N}},$ and are diagonal\\ matrices. The realisability condition  in  $P2$ forces $\Pi^{*}_{1}$ and $\Pi^{*}_{0}$ to be diagonal matrices. Let $\Pi^{*}_{1} = \text{diag}(n^{1}_{1},\ldots, n^{1}_{N})$ and $\Pi^{*}_{0} = \text{diag}(1-n^{1}_{1},\ldots, 1- n^{1}_{N})$. Then for optimality, $\zeta_{0}p^{0}_{i}n^{1}_{i} + \zeta_{1}p^{1}_{i}(1 - n^{1}_{i} ) \leq \zeta_{0}p^{0}_{i}$ and \\ $\zeta_{0}p^{0}_{i}n^{1}_{i} + \zeta_{1}p^{1}_{i}(1 - n^{1}_{i} ) \leq \zeta_{1}p^{1}_{i},  1  \leq i \leq N$. For both inequalities to hold, it follows that if $\zeta_{0}p^{0}_{i} \geq \zeta_{1}p^{1}_{i}$, then $n^{1}_{i} = 0$. Else $n^{1}_{i} = 1$. 

\noindent The minimum cost is, $\mathbb{P}^{*}_{e} = \sum^{N}_{i=1}\min\{\zeta_{0}p^{0}_{i},  \zeta_{1}p^{1}_{i}\} \leq \min\{\zeta_{0},  \zeta_{1}\}$. Clearly \\ $\alpha^{j}_{i}= n^{j}_{i}, \; 1\leq i \leq N,j =1, 0$. As in the classical probability scenario, we \\ obtain pure strategies, i.e , when measurement $i$ is obtained , if $\zeta_{0}p^{0}_{i} \geq \zeta_{1}p^{1}_{i}$ then the decision is $0$ with probability $1$, else decision is $1$ with probability $1$.  For a discussion on why the minimum probability of error does not depend on the choice of (state, PVM) representation we refer to \cite{raghavan2019non}.

\subsubsection{Positive operator valued measure}\label{Binary hypothesis testing problem Solution Positive operator valued measure}
To find the states and the POVM, a numerical method is proposed. If a feasibility problem is formulated with the state and POVM as optimization variables, the resulting problem is nonconvex. Hence we consider a finite set of states, $\mathcal{S} \subset  \mathcal{T}^{+}_{s}(\mathbb{C}^{k}) $, $|\mathcal{S}| < \infty$. For a pair of states, $(\hat{\rho}_{0}, \hat{\rho}_{1}) \in \mathcal{S} \times \mathcal{S}$, $\hat{\rho}_{0} \neq \hat{\rho}_{1}$, the following feasibility problem must be solved:
\begin{align*}
&P6: \underset{t \in \mathbb{R},\{ M(i)\}^{N}_{i=1} \subset \mathbb{C}^{k \times k} }  \min t \;\; \text{s.t}  \; Tr[\hat{\rho}_{h}M(i)] - p^{h}_{i} = t, h =0,1 , 1 \leq i \leq N\\\noalign{\vskip-10pt}
& M(i) \leq -t , 1 \leq i \leq N, \sum^{N}_{i=1} M(i) - I_{\mathbb{C}^{k}} =tI_{\mathbb{C}^{k}}.
\end{align*} 
If for a particular pair of states, $\hat{\rho}_{0}, \hat{\rho}_{1}$ the optimal value of the above feasibility problem, $t^*$ is less than or equal to zero, then the corresponding minimizers $\{M(i)\}^{N}_{i=1}$ form the POVM. If for every pair of states, the optimal value of the feasibility problem is greater than zero, then optimization problems need to be solved for a new set of states. In Appendix \ref{Existence of a State for a Given P.O.V.M}, section \ref{Existence of a State for a Given P.O.V.M Problem considered}, we consider the problem where given a POVM  and a finite dimensional probability distribution, we need to check if there exists a state such that the state and POVM combination achieves the probability distribution. In Appendix \ref{Existence of a State for a Given P.O.V.M}, section \ref{Existence of a State for a Given P.O.V.M Solution}, we present sufficient conditions under which this feasibility problem can be solved. 
\vspace{-\baselineskip}
\begin{lemma}\label{Binary hypothesis testing problem Solution Positive operator valued measure Lemma 1}
Suppose $\bar{\mathbb{S}} \neq \emptyset$. Let $\mathbb{Q}^{*}_{e} =  \underset{(\hat{\rho}_{0}, \hat{\rho}_{1},M) \in \bar{\mathbb{S}}} \min Q(\hat{\rho}_{0}, \hat{\rho}_{1},M).$ Then, \\ $\mathbb{Q}^{*}_{e} = \mathbb{P}^{*}_{e}$.
\end{lemma}
\vspace{-\baselineskip}
\noindent For the proof of the Lemma, solution to problem $P5$, and discussion on the above Lemma we refer to \cite{raghavan2019non}.
\subsection{Numerical results}
Consider the scenario described in the beginning of section \ref{Binary hypothesis testing problem Problem formulation Positive operator valued measure}. We describe a simple example of that scenario. Let $\eta_{1} = 3$ and $\eta_{2} = 2$. When $Y_{2}$ is collected after $Y_{1}$,  the distribution of the observations under hypothesis $0$ and $1$ is tabulated in the second and third columns of Table \ref{Table 4} respectively. When $Y_{1}$ is collected after $Y_{2}$,  the distribution of the observations under hypothesis $0$ and $1$ is tabulated in the fifth and sixth columns of Table \ref{Table 4} respectively. The prior distribution of the hypothesis is set to $(\zeta_{0} =0.4,\zeta_{1} =0.6 )$. The minimum probability of error is calculated using Lemma \ref{Binary hypothesis testing problem Solution Positive operator valued measure Lemma 1}. The minimum probability of error when $Y_{2}$ is measured after $Y_{1}$ is $0.35$. The minimum probability of error when $Y_{1}$ is measured after $Y_{2}$ is $0.266$. Hence in this example the optimal strategy is first measure $Y_{2}$ and then measure $Y_{1}$.  
\begin{table}[h!]
\begin{center}
\begin{tabular}{||c || c|| c ||c ||c ||c ||} 
 \hline
 [$Y_{1}, Y_{2}$]  & $h=0$ & $h=1$ &[$Y_{2}, Y_{1}$]& $h=0$ &$h=1$ \\
 \hline
 $1,1$& $0.1$ &$0.15$ & $1,1$ & $0.25$ &$0.15$\\
 \hline
 $1,2$ & $0.2$ &$0.3$ & $2,1$ & $0.05$ &$0.30$ \\
 \hline
 $2,1$ & $0.2$ &$0.15$ &$1,2$ & $0.25$ &$0.13$ \\
 \hline
 $2,2$ & $0.15$ &$0.25$ &$2,2$ & $0.1$ &$0.27$  \\
 \hline
 $3,1$ & $0.25$ &$0.1$  &$1,3$ & $0.05$ &$0.12$\\
 \hline
 $3,2$ & $0.1$ &$ 0.05$ &$2,3$ & $0.3$ &$ 0.03$\\
 \hline
\end{tabular}
\vspace{0.2cm}
\caption{Distribution of observations under either hypothesis }
\label{Table 4}
\end{center}
\vspace{-\baselineskip}
\vspace{-\baselineskip}
\end{table}
Next, We consider the problem described in section \ref{Example: multi-agent decision making Problem description} and the marginal distributions mentioned in Table \ref{Table 1}. The state $\rho$, the projections $E_{1}$, $E_{2}$, $E_{3}$ which achieve the empirical distributions of the decisions, are not necessarily unique. The set $\{ \rho \in \mathbb{S}, E_{1}, E_{2}, E_{3} \in \mathcal{P}(\mathbb{R}^2) : Tr[\rho E_{i}] = \mathbb{P}(D_{i} =1), i =1,2,3 \}$ is not necessarily a singleton set. Given the ordered distributions, distribution of $D_{1}$ and then $D_{2}$, distribution of $D_{1}$ and then $D_{3}$, etc., it might be possible to find $\rho, E_{1}, E_{2}, E_{3}$ uniquely. There are six different orders in which measurements can be collected. Given unique $E_{1}$, $E_{2}$ and $E_{3}$, the POVM for each order measurement can be found uniquely. This problem has not been addressed in this paper. We directly consider a POVM representation for each order of measurement. The hypothesis testing problem for the central coordinator is formulated as in section \ref{Binary hypothesis testing problem Problem formulation Positive operator valued measure} and solved as in \ref{Binary hypothesis testing problem Solution Positive operator valued measure}. The minimum probability of error that can be achieved for each order of measurement is calculated using lemma \ref{Binary hypothesis testing problem Solution Positive operator valued measure Lemma 1} and tabulated in Table \ref{Table 5}. The sequence of measurements where $D_{i}$ is measured first, followed by $D_{j}$, and then $D_{k}$ is denoted as $D_{i}, D_{j}, D_{k}$. The two orders in which $D_{3}$ is measured first,  $D_{3}, D_{1}, D_{2}$ and $D_{3}, D_{2}, D_{1}$ have higher probability of error. 
\begin{table}[h!]
\vspace{-\baselineskip}
\begin{center}
\begin{tabular}{||c || c||} 
 \hline
 Order of measurements  & Probability of error \\
 \hline
 $D_{2}, D_{1}, D_{3}$& $0.1740$\\
 \hline
 $D_{1}, D_{2}, D_{3}$& $0.1713$\\
 \hline
 $D_{3}, D_{1}, D_{2}$& $0.1913$\\
 \hline
 $D_{1}, D_{3}, D_{2}$& $0.1711$\\
 \hline
 $D_{2}, D_{3}, D_{1}$& $0.1745$\\
 \hline
 $D_{3}, D_{2}, D_{1}$& $0.1918$\\
 \hline
\end{tabular}
\vspace{0.2cm}
\caption{Minimum probability of error for different orders of measurements}
\label{Table 5}
\end{center}
\vspace{-\baselineskip}
\vspace{-\baselineskip}
\end{table}
\section{Conclusion}\label{Conclusion}
To conclude, in the first section of this paper we discussed a methodology from the literature which can be used to investigate the structure of the set of events. In the second section, we considered a multi-agent hypothesis testing problem with three observers and a central coordinator. The structure of the set of events for the central coordinator was studied. We showed that the set of events did not form a Boolean algebra, instead form an ortholattice. In the third section we considered the binary hypothesis testing problem with finite observation space. First, the measurements were represented using PVM, and  the detection problem was formulated to minimize the probability of error. The solution to the detection problem was pure strategies and the expected cost with optimal strategies was the same as the minimum probability of error that could be achieved using classical probability models. In another approach, the measurements were represented using POVM and the hypothesis testing problem was solved. This approach was used for the central coordinator in the multi-agent hypothesis testing problem resulting in different minimum probabilities of error for different orders of measurement.  
\vspace{-\baselineskip}
\section{Appendix}\label{Verification of axioms II.4 and II.5}
\vspace{-\baselineskip}
\subsection{Verification of Axiom II.4}\label{Verification of axioms II.4 and II.5 Axiom II.4}
Let the domain of $T = T_{E_{1}} \circ T_{E_{2}} \circ \ldots \circ T_{E_{n}} = T_{F_{1}} \circ T_{F_{2}} \circ \ldots \circ T_{F_{n}}$ be $\mathbb{D}_{T}$. Let $U = \prod^{1}_{i=n}E_{i} = E_{n}E_{n-1}\ldots E_{1}$, $ U^{\ast}= \prod^{n}_{i=1}E_{i} = E_{1}E_{2}\ldots E_{n}$,\\ $V = \prod^{1}_{i=n}F_{i}$ and $V^{\ast} = \prod^{n}_{i=1}F_{i}$. Thus $\mathbb{D}_{T} = \{\rho \in \mathbb{S} : Tr[U^{\ast}\rho U] \neq 0 \} = \{\rho \in \mathbb{S} : Tr[V^{\ast}\rho V] \neq 0\}$. $T_{E_{1}} \circ T_{E_{2}} \circ \ldots \circ T_{E_{n}} =  T_{F_{1}} \circ T_{F_{2}} \circ \ldots \circ T_{F_{n}}$ is equivalent to $\frac{U^{\ast}\rho U}{Tr[U^{\ast}\rho U]} = \frac{V^{\ast}\rho V}{Tr[V^{\ast}\rho V]} \; \forall \rho \in \mathbb{D}_{T}$. We claim that $\exists \alpha \in \mathbb{C}, \alpha \neq 0$ such that $U = \alpha V$. We prove by contradiction. Suppose our claim is not true. Then for every $\alpha$, there exists $h_{1} \in \mathcal{H}, h_{1} \neq \theta$ and $h_{2} \in \mathcal{H}, h_{2} \neq \theta$ such that $U(h_{1}) \neq \alpha V(h_{1})$ and $U^{\ast}(h_{2}) \neq \bar{\alpha} V^{\ast}(h_{2})$ where $\bar{\alpha}$ denotes the complex conjugate of $\alpha$. Let $\rho(h) = \frac{\langle h, h_{2}\rangle  h_{2}}{|| h_{2}||^{2}} \; \forall h \in \mathcal{H}$. Hence $\rho $ is the orthogonal projection on to the subspace spanned by $h_{2}$. $\langle\rho(h),h\rangle \geq 0 \; \forall h \in \; \mathcal{H}$ and $\rho = \rho^{\ast}$. $Tr[\rho] = \sum_{i}\langle\rho(e_{i}),e_{i}\rangle \\ = \frac{1}{||h_{2}||^{2}}\sum_{i}\langle h_{2},e_{i}\rangle^2 =1$. Hence $\rho \in \mathbb{S}$.

\noindent \underline{Case 1}: Suppose $h_{2}$ is such that $h_{2} \in \mathcal{N}(U^{\ast})$. Then $h_{2} \notin \mathcal{N}(\bar{\alpha} V^{\ast})$. $h_{2} \in \mathcal{N}(U^{\ast})$ implies that $Tr[U^{\ast}\rho U] = 0$. $\langle \alpha V(h), h_{2}\rangle = 0 \;  \forall h \in \mathcal{H}$ implies that $h_{2} \\ \perp \mathcal{R}(\alpha V)$. Since $[\mathcal{R}(\alpha V)]^{\perp} = \mathcal{N}(\bar{\alpha} V^{\ast})$, it follows that $h_{2} \in \mathcal{N}(\bar{\alpha} V^{\ast})$. Hence, $h_{2} \notin \mathcal{N}(\bar{\alpha} V^{\ast})$ implies that $\exists h_{3} \in \mathcal{H}$ such that $\langle \alpha V(h_{3}),h_{2}\rangle \neq 0$. This further implies that $Tr[V^{\ast}\rho V] \neq 0 $. Hence the domains of the two operations $T_{E_{1}} \circ T_{E_{2}} \circ \ldots \circ T_{E_{n}}$ and $T_{F_{1}} \circ T_{F_{2}} \circ \ldots \circ T_{F_{n}}$ are unequal which implies that the operations are unequal. Similarly, we can obtain a contradiction if  $h_{2} \in \mathcal{N}(\bar{\alpha} V^{\ast})$ and $h_{2} \notin \mathcal{N}(U^{\ast})$.

\noindent \underline{Case 2}: Let $U(h_{1}) = \alpha V(h_{1}) + h_{3}$. $\rho(U(h_{1})) = \frac{\langle U(h_{1}) , h_{2}\rangle  h_{2}}{|| h_{2}||^{2}} $ and $\rho(\alpha V(h_{1})) = \frac{  \langle \alpha V(h_{1}) , h_{2}\rangle  h_{2}}{|| h_{2}||^{2}} $.
\begin{align*}
U^{\ast}(\rho(U(h_{1}))) = \frac{\langle U(h_{1}) , h_{2}\rangle}{|| h_{2}||^{2}}U^{\ast}( h_{2})\; \text{and} \\ \bar{\alpha}V^{\ast}(\rho(\alpha V(h_{1}))) = \frac{\langle \alpha V(h_{1}) , h_{2}\rangle}{|| h_{2}||^{2}}\bar{\alpha}V^{\ast}(h_{2})
\end{align*}
$\langle U^{\ast}(u),v \rangle = \langle u,U(v)\rangle \forall u,v \in \mathcal{H}$. Letting $u = U(h_{1}), v = h_{1}$ we get, 
\begin{align*}
\langle U^{\ast}(U(h_{1})), h_{1} \rangle &= \langle U(h_{1}),U(h_{1})\rangle \\
&= \langle U(h_{1}),\alpha V(h_{1})\rangle + \langle U(h_{1}), h_{3}\rangle\\
&= \langle \bar{\alpha}V^{\ast}U(h_{1}),h_{1}\rangle + \langle \alpha V(h_{1}) + h_{3},h_{3}\rangle. \\
\langle U^{\ast}(U(h_{1}))- &\bar{\alpha}V^{\ast}(U(h_{1})),  h_{1} \rangle =\langle \alpha V(h_{1}),h_{3}\rangle + \langle h_{3},h_{3}\rangle. 
\end{align*}
Suppose $U(h_{1}) \neq \theta$ and $\langle \alpha V(h_{1}),h_{3}\rangle + \langle h_{3},h_{3}\rangle \neq 0$. \\ Then $U^{\ast}(U(h_{1})) \neq \bar{\alpha}V^{\ast}(U(h_{1}))$. We let $h_{2} = U(h_{1})$. For this choice of $h_{2}$, if $h_{2} \in \mathcal{N}(U^{\ast})$(and $h_{2} \notin \mathcal{N}(\bar{\alpha} V^{\ast})$) or $h_{2} \in \mathcal{N}(\bar{\alpha} V^{\ast})$(and $h_{2} \notin \mathcal{N}(U^{\ast}))$ then we already have a contradiction (by case 1). We consider the scenario where $h_{2} \notin \mathcal{N}(U^{\ast})$ and $h_{2} \notin \mathcal{N}(\bar{\alpha} V^{\ast})$. Thus, $\rho$ belongs to the domain of both operations. 
\begin{align*}
&U^{\ast}(\rho(U(h_{1}))) = U^{\ast}(U(h_{1})) \; \text{and}\\
&\bar{\alpha}V^{\ast}(\rho(\alpha V(h_{1}))) = \frac{\langle \alpha V(h_{1}) , U(h_{1})\rangle}{||U(h_{1}) ||^{2}}\bar{\alpha}V^{\ast}(U(h_{1})).\\
&U^{\ast}(\rho(U(h_{1}))) = \bar{\alpha}V^{\ast}(\rho(\alpha V(h_{1}))) \iff \exists \beta \neq 1: \\ & U^{\ast}(U(h_{1})) = \beta \bar{\alpha}V^{\ast}(U(h_{1})) \; \text{and} \; \beta = \frac{\langle \alpha V(h_{1}) , U(h_{1})\rangle}{||U(h_{1}) ||^{2}}. \\
&\beta =  \frac{\langle \alpha V(h_{1}) , U(h_{1})\rangle}{||U(h_{1}) ||^{2}} \Longleftrightarrow \beta = \frac{\langle h_{1} ,\bar{\alpha}V^{\ast}( U(h_{1}))\rangle}{||U(h_{1}) ||^{2}}\Longleftrightarrow \\ &\beta = \frac{1}{\bar{\beta}}\frac{\langle h_{1} ,U^{\ast}( U(h_{1}))\rangle}{||U(h_{1}) ||^{2}} \Longleftrightarrow \beta = \frac{1}{\bar{\beta}}  \Longleftrightarrow |\beta|^{2} = 1.
\end{align*}
$\beta \neq -1$ as $U^{\ast} \rho U \geq 0, V^{\ast} \rho V \geq 0$. Hence $U^{\ast}(\rho(U(h_{1}))) \neq \bar{\alpha}V^{\ast}(\rho(\alpha V(h_{1})))$. The above proof holds even if $\alpha V(h_{1}) = \theta$. 

\noindent \underline{Case 3}: Suppose $\langle \alpha V(h_{1}),h_{3}\rangle + \langle h_{3},h_{3}\rangle = 0$ ($U(h_{1}) = \theta$  implies the same). $\langle \bar{\alpha}V^{\ast}(u),v \rangle = \langle u,\alpha V(v)\rangle \forall u,v \in \mathcal{H}$. Letting $u =  \alpha V(h_{1}), v=h_{1}$ we get,
\begin{align*}
\langle \bar{\alpha}V^{\ast}(\alpha V(h_{1} ))),h_{1} \rangle &= \langle \alpha V(h_{1}),\alpha V(h_{1})\rangle\\
&= \langle \alpha V(h_{1}),U(h_{1})\rangle - \langle \alpha V(h_{1}),h_{3}\rangle \\
&= \langle U^{\ast}(\alpha V(h_{1})),h_{1}\rangle + \langle h_{3},h_{3}\rangle. \\
\langle \bar{\alpha}V^{\ast}(\alpha V(h_{1})))  - &U^{\ast}(\alpha V(h_{1})) ,h_{1} \rangle  = \langle h_{3},h_{3}\rangle \neq 0.
\end{align*} 
Hence $U^{\ast}(\alpha V(h_{1}))\neq \bar{\alpha}V^{\ast}(\alpha V(h_{1}))) $. Let $h_{2} = \alpha V(h_{1})$. For this choice of $h_{2}$, if $h_{2} \in \mathcal{N}(U^{\ast})$(and $h_{2} \notin \mathcal{N}(\bar{\alpha} V^{\ast})$) or $h_{2} \in \mathcal{N}(\bar{\alpha} V^{\ast})$(and $h_{2} \notin \mathcal{N}(U^{\ast}))$ then we already have a contradiction (by case 1). We consider the scenario where $h_{2} \notin \mathcal{N}(U^{\ast})$ and $h_{2} \notin \mathcal{N}(\bar{\alpha} V^{\ast})$. Thus, $\rho$ belongs to the domain of both operations. Using the definition of $\rho$,
\begin{align*}
&U^{\ast}(\rho(U(h_{1}))) = \frac{\langle U(h_{1}) , \alpha V(h_{1})\rangle}{|| \alpha V(h_{1})||^{2}}U^{\ast}(\alpha V(h_{1})) \; \text{and} \\ 
&\bar{\alpha}V^{\ast}(\rho(\alpha V(h_{1}))) = \bar{\alpha}V^{\ast}(\alpha V(h_{1})).\\
&U^{\ast}(\rho(U(h_{1}))) = \bar{\alpha}V^{\ast}(\rho(\alpha V(h_{1}))) \iff \exists \beta \neq 1 :\\
&\beta U^{\ast}(\alpha V(h_{1})) = \bar{\alpha}V^{\ast}(\alpha V(h_{1}))  \; \text{and} \;\beta = \frac{\langle U(h_{1}) , \alpha V(h_{1}))\rangle}{||U(h_{1}) ||^{2}}.\\
&\beta = \frac{\langle U(h_{1}) , \alpha V(h_{1})\rangle}{|| \alpha V(h_{1}) ||^{2}} \Longleftrightarrow \beta = \frac{\langle h_{1} , U^{\ast}(\alpha V(h_{1}))\rangle}{|| \alpha V(h_{1}) ||^{2}}\Longleftrightarrow \\
&\beta = \frac{1}{\bar{\beta}}\frac{\langle h_{1} ,\bar{\alpha}V^{\ast}(\alpha V(h_{1})) \rangle}{|| \alpha V(h_{1}) ||^{2}} \Longleftrightarrow |\beta|^{2} = 1.
\end{align*} 
Again, $\beta \neq -1$. Hence $U^{\ast}(\rho(U(h_{1}))) \neq \bar{\alpha}V^{\ast}(\rho(\alpha V(h_{1})))$. \\ For every $\alpha$ in $\mathbb{C}$, $\exists \rho \in \mathbb{S}$, $U^{*}\rho U \neq |\alpha|^{2}V^{*}\rho V$. Thus, $\frac{U^{*}\rho U}{Tr[U^{*}\rho U]} \neq \frac{V^{*}\rho V}{Tr[V^{*}\rho V]}$, if not for $\alpha_{1}$, such that $|\alpha_{1}|^{2} \frac{Tr[U^{*}\rho U]}{Tr[V^{*}\rho V]}$, there should not exist $\rho \in \mathbb{D}_{T}$ such that $U^{*}\rho U \neq |\alpha_{1}|^{2}V^{*}\rho V$. This a contradiction. 

\noindent Hence our claim is true. Since $U = \alpha V$, $U^{\ast} = \bar{\alpha} V^{\ast}$. Hence for every $\rho \in \mathbb{S}$, $\rho U^{\ast} = \rho \bar{\alpha} V^{\ast}$. This implies that $U \rho U^{\ast} = |\alpha|^{2} V \rho  V^{\ast} \; \forall \rho \in \mathbb{D}_{T}$, which further implies that $\frac{U \rho U^{\ast}}{Tr[U \rho U^{\ast}]} = \frac{V \rho  V^{\ast}}{Tr[V \rho  V^{\ast}]} \forall \rho \in \mathbb{D}_{T}$. The final equality is equivalent stating that $T_{E_{n}} \circ T_{E_{n-1}} \circ \ldots T_{E_{1}} = T_{F_{n}} \circ T_{F_{n-1}} \circ \ldots T_{F_{1}}$, thus verifying Axiom [II.4]. 
\subsection{Verification of Axiom II.5}\label{Verification of axioms II.4 and II.5 Axiom II.5}
For $T \in \mathbb{O}_{\mathbb{T}}$, there exists $E_{1}, E_{2}, \ldots, E_{n}$, such that $T = T_{E_{1}} \circ T_{E_{2}} \circ \ldots T_{E_{n}} = \frac{\prod^{n}_{i=1}E_{i}\rho \prod^{1}_{i=n} E_{i} }{Tr[\prod^{n}_{i=1}E_{i}\rho \prod^{1}_{i=n} E_{i}]}$. The domain of $T$ is $\mathbb{D}_{T}=\{ \rho \in \mathbb{S}: Tr[\prod^{n}_{i=1}E_{i}\rho \prod^{1}_{i=n}E_{i}]\\ \neq 0\}$. The set of states which do not belong to the domain is \\$\{\rho \in \mathbb{S}: Tr[\prod^{n}_{i=1}E_{i}\rho \prod^{1}_{i=n}E_{i}] = 0\}$.  
\begin{align*}
\{\rho \in \mathbb{S}: Tr[\prod^{n}_{i=1}E_{i}\rho \prod^{1}_{i=n} E_{i}] = 0\} 
&\overset{(a)}{=} \{\rho \in \mathbb{S}: \prod^{n}_{i=1}E_{i}\rho \prod^{1}_{i=n}E_{i} = \Theta\} \\ &= \{\rho \in \mathbb{S}: \mathcal{R}(\rho \prod^{1}_{i=n}E_{i}) \subseteq \mathcal{N}(\prod^{n}_{i=1}E_{i})\}. 
\end{align*}
The equality $ \overset{(a)}{=}$ follows from the observation that $\prod^{n}_{i=1}E_{i}\rho \prod^{1}_{i=n}E_{i}$ is a positive semi-definite operator. Let $Q$ denote the orthogonal projection on to the null space of $\prod^{n}_{i=1}E_{i}$, $\mathcal{N}(\prod^{n}_{i=1}E_{i})$, which is a closed subspace. $\mathcal{N}(\prod^{n}_{i=1}E_{i}) = \mathcal{R}(Q)$. Then, 
\begin{align*}
\{\rho \in \mathbb{S}:\mathcal{R}(\rho \prod^{1}_{i=n}E_{i}) \subseteq \mathcal{N}(\prod^{n}_{i=1}E_{i})\}
&\overset{(b)}{=} \{\rho \in \mathbb{S}: \rho (\mathcal{R}(I-Q))\subset \mathcal{R}(Q)\}\\
& \overset{(c)}{=} \{\rho \in \mathbb{S}: (I-Q)\rho(I-Q) = \Theta\} \\
&= \{\rho: Tr[\rho Q] =1\}
\end{align*}
The equality, $ \overset{(b)}{=}$, can be proven as follows. First we note that,\\  $[\mathcal{R}(Q)]^{\perp} = \mathcal{R}(I-Q)$. The closure of the range of $\prod^{1}_{i=n}E_{i}$ is the the closed subspace $\mathcal{R}(I-Q)$. For all $h \in \overline{\mathcal{R}(\prod^{1}_{i=n}E_{i})}$, $\rho(h) \in \mathcal{R}(Q)$ implies that \\ $\rho(h) \in \mathcal{R}(Q)\; \forall h \in \mathcal{R}(\prod^{1}_{i=n}E_{i})$. Thus, 
\begin{align*}
\{\rho \in \mathbb{S}: \rho (\mathcal{R}(I-Q))\subset \mathcal{R}(Q)\} \subseteq \{\rho \in \mathbb{S}:\mathcal{R}(\rho \prod^{1}_{i=n}E_{i}) \subseteq \mathcal{N}(\prod^{n}_{i=1}E_{i})\}.  
\end{align*}
Let $h$ be a closure point of $\mathcal{R}(\prod^{1}_{i=n}E_{i})$, i.e., $\exists \{h_{n}\}_{n \geq 1} \subset \mathcal{R}(\prod^{1}_{i=n}E_{i}) \; \text{s.t} \\ \; \{h_{n}\} \rightarrow h, h \notin  \mathcal{R}(\prod^{1}_{i=n}E_{i})$. By continuity of $\rho$, $\{\rho(h_{n})\} \rightarrow \rho(h)$. Since $\mathcal{R}(Q)$ is closed, $\rho(h_{n}) \in \mathcal{R}(Q) \; \forall n$ implies that $\rho(h) \in \mathcal{R}(Q)$. Hence, 
\begin{align*}
\{\rho \in \mathbb{S}:\mathcal{R}(\rho \prod^{1}_{i=n}E_{i}) \subseteq \mathcal{N}(\prod^{n}_{i=1}E_{i})\} \subseteq \{\rho \in \mathbb{S}: \rho (\mathcal{R}(I-Q))\subset \mathcal{R}(Q)\}.  
\end{align*}
The equality, $ \overset{(c)}{=}$, can be proven as follows. Suppose $\rho$ is such that $ \rho (\mathcal{R}(I-Q))\subset \mathcal{R}(Q)$.  Then for every $h \in \mathcal{H}$, $\rho(I-Q)h \in \mathcal{R}(Q)$, which implies that $(I-Q)\rho(I-Q)h = \Theta$ as $\mathcal{R}(Q)$ is the null space of $I-Q$. Hence,
\begin{align*}
\{\rho \in \mathbb{S}: \rho (\mathcal{R}(I-Q)) \subset \mathcal{R}(Q)\} \subseteq \{\rho \in \mathbb{S}: (I-Q)\rho(I-Q) = \Theta\}.
\end{align*}
Suppose $\rho$ is such that $(I-Q)\rho(I-Q) = \Theta$. Then $\mathcal{R}(\rho(I-Q))$ is subset of the null space of $I-Q$ which is $\mathcal{R}(Q)$. Thus, 
\begin{align*}
\{\rho \in \mathbb{S}: (I-Q)\rho(I-Q) = \Theta\} \subset \{\rho \in \mathbb{S}: \rho ((I-Q)\mathcal{H})\subseteq Q(\mathcal{H})\},
\end{align*} proving that the two sets are indeed equal. Hence, there exists $Q \in \mathcal{P}(\mathcal{H})$ such that $\{\rho \in \mathbb{S}: Tr[\prod^{n}_{i=1}E_{i}\rho \prod^{1}_{i=n}E_{i}]= 0\} = \{\rho \in \mathbb{S}: Tr[\rho Q] =1\}$, verifying Axiom II.5. 
\vspace{-\baselineskip}
\subsection{Existence of a State for a Given P.O.V.M}\label{Existence of a State for a Given P.O.V.M}
\vspace{-\baselineskip}
\subsubsection{Problem considered}\label{Existence of a State for a Given P.O.V.M Problem considered}
Let $\{p_{i}\}_{\{1 \leq i \leq N\}}$ be a given probability distribution on a finite observation space. Let $X = \{1,2,...,N\}$, $\Sigma  = 2^{X}$. Let $O$ be a POVM from $\Sigma$ on to $\mathcal{B}^{+}_{s}(\mathbb{C}^{k})$, where $\mathcal{B}^{+}_{s}(\mathbb{C}^{k})$ denotes the set of positive semidefinite Hermitian matrices on $\mathbb{C}^{k}$. The objective is to find sufficient conditions on $O$, so that $\exists \rho$, $\rho \in  \mathcal{T}^{+}_{s}(\mathbb{C}^{k})$ such that, $Tr[\rho O(i)] = p^{h}_{i}, 1 \leq i \leq N,$ where $Tr[\cdot]$ is the trace operator. 
\subsubsection{Solution}\label{Existence of a State for a Given P.O.V.M Solution}
Let $M_{k}$ be the vector space of $k \times k$ complex matrices over the field of real numbers. The dimension of $M_{k}$ is $2k^{2}$. Let $H_{k}$ be the subspace of hermitian matrices. The dimension of $H_{k} $is $k^{2}$. Let $S_{k}$ be the cone of positive semi-definite matrices. $S_{k}$ is closed and convex. Let the vector space be endowed with following inner product,$\langle A,B \rangle = Tr [A^{H}B],$ where $A^{H}$ denotes the conjugate transpose of the matrix $A$. Let $\{e_{mn}\}_{\{1 \leq m,n \leq k \} }$ be a set of orthonormal basis vectors for the subspace $H_{k}$. For every matrix $O \in S_{k}$, there exits unique real numbers $O_{mn}$ such that $O = \sum_{\{1 \leq m,n \leq k \}}O_{mn}e_{mn}$. The $k^{2}$ dimensional vector obtained from the real numbers is represented by $\bar{O}$. Let the collection of all the vectors obtained from the matrices in $S_{k}$ be represented by $\bar{S}_{k}$. $\bar{S}_{k}$ is a closed convex cone in $\mathbb{R}^{k^{2}}$. Hence, for each $O(i)$, there exists unique real numbers $O_{mn}(i)$ such that, $O(i) = \sum_{\{1 \leq m,n \leq k \}}O_{mn}(i)e_{mn}$ and the corresponding vectors are represented by $\bar{O}_{i}$. $P= [p_{1}; p_{2},\ldots,p_{N}]$ is a $N$ dimensional vector. Let $A = [\bar{O}^{H}_{1},\bar{O}^{H}_{2}, \ldots, \bar{O}^{H}_{N}]$. Let $C = \{Ax, x \in \bar{S}_{k}\}$. From \cite{berman1973linear}, we note that $\bar{S}_{k}$ is self-dual cone. Hence the original problem can be recast as: Is $P \in C$ or $P \not\in C $. It should be noted that $C$ is a convex cone and is not necessarily closed. One of the sufficient conditions for $C$ to be closed is mentioned in \cite{pataki2007closedness}. The condition is that $\text{ri}(\bar{S}_{k}) \cap \mathcal{R}(A^{H}) \neq \emptyset$. $\text{ri}(S)$ denotes the relative interior of a set $S$ and is defined as $\text{ri}(S)= \{ x \in S: \exists \epsilon >0 , N_{\epsilon}(x) \cap \text{aff}(S) \subseteq S\}$, where $\text{aff}(S)$ denotes the affine hull of $S$. The affine hull of $S_{k}$ is $H_{k}$. The positive definite matrices belong to the interior of $S_{k}$. Also $\mathcal{R}(A^{H}) = \text{span}(\bar{O}_{1},\bar{O}_{2}, \ldots, \bar{O}_{N} )$. Hence, if one element of the POVM is a positive definite matrix then the sufficient condition for the closedness of the set $C$ is satisfied. The first condition imposed on the POVM is that at least one of the elements is positive definite. With this condition, set $C$ is closed convex cone and set $\{P\}$ is closed, convex and compact. Hence if the two sets are disjoint, i.e, $\nexists x \in \bar{S}_{k}: Ax = P$, then by separating hyperplane theorem there exists a vector $v$ and real number $\alpha >0$ such that,$v^{H}P < \alpha \;\; \text{and} \; v^{H}c > \alpha \;\forall \; c \in C.$ Since $C$ is cone it follows that,$ v^{H}P < 0 \;\; \text{and} \; v^{H}c > 0 \; \forall \; c \in C.$ Hence, if there does not exist a vector $v$ such that $v^{H}Ax > 0 \; \forall x \in \bar{S}_{k}$ and $v^{H}P <0$, then the state $\rho$ exists. 
\vspace{-\baselineskip}

\address{Institute of Systems Research, University of Maryland\\
College Park, MD, 20740, USA\\
\email{raghava@umd.edu}}

\address{Institute of Systems Research, University of Maryland\\
College Park, MD, 20740, USA\\
\email{baras@umd.edu}}
\end{document}